\documentclass[showpacs,preprintnumbers,prd,amsmath,amssymb]{revtex4}
\usepackage{amsmath}
\usepackage{graphicx}
\usepackage{subfig}
\usepackage{hyperref}
\usepackage{amssymb}
\usepackage[latin1]{inputenc}
\usepackage[english]{babel}
\usepackage{epsfig}
\usepackage{wasysym}

\begin{document}

\title{Regular multihorizon black holes in $f(G)$ gravity with nonlinear electrodynamics\\}

\author{Manuel E. Rodrigues$^{(1,2)}$\footnote{E-mail address:esialg@gmail.com}, Marcos V. de S. Silva$^{(1)}$\footnote{E-mail address:marco2s303@gmail.com}}

\affiliation{$^{(1)}$Faculdade de Física, Programa de Pós-Graduação em Física, Universidade Federal do Pará, 66075-110, Belém, Pará, Brazil\\
	$^{(2)}$Faculdade de Ci\^{e}ncias Exatas e Tecnologia, Universidade Federal do Par\'{a} Campus Universit\'{a}rio de Abaetetuba, 68440-000, Abaetetuba, Par\'{a}, 
	Brazil
}


\begin{abstract}
In this work, we study the existence of regular black holes solutions with multihorizons in general relativity and in some alternative theories of gravity. We consider the coupling between the gravitational theory and nonlinear electrodynamics. The coupling generates modifications in the electromagnetic sector. This paper has as main objective generalize solutions already known from general relativity to the $f(G)$ theory. To do that, we first correct some misprints of the Odintsov and Nojiri's work in order to introduce the formalism that will be used in the $f(G)$ gravity. In order to satisfy all field equations, the method to find solutions in alternative theories generates different $f(R)$ and $f(G)$ functions for each solution, where only the nonlinear term of $f(G)$ contributes to the field equations. We also analyze the energy conditions, since it is expected that some must be violated to find regular black holes, and using an auxiliary field, we analyze the nonlinearity of the electromagnetic theory.
\end{abstract}

\date{\today}

\maketitle



\section{Introduction}\label{sec1}
Black holes are one of the most interesting predictions of general relativity \cite{wal}. These objects have a region of non-scape where the boundary is a surface which permits the passage only in one direction, the event horizon \cite{visser}. The most simple black hole solution is described by the Schwarzschild metric, which is characterized only by its mass \cite{chan}. The structure of the Schwarzschild black hole is composed by an event horizon and a singularity in the black hole center \cite{din}. There are several metrics that are more general than the Schwarzschild solution, such as the Reissner-Nordström (electrically charged), Kerr (with rotation) or de Sitter-like solutions (cosmological constant) \cite{hans}. The presence of these other parameters can lead to changes in the structure of black holes such as the Cauchy and cosmological horizon \cite{chan,hans}.

Although some solutions have a singularity, this characteristic is not necessary for black holes. Actually, it is possible to find solutions that have an event horizon without singularity; this kind of solution is known as regular black hole \cite{Stefano}. James Bardeen proposed a metric which was later interpreted as a solution to the Einstein equations for nonlinear electrodynamics \cite{Beato1}, without the presence of singularities \cite{Bardeen}. As in the Reissner-Nordström case, due to the charge, Bardeen solution presents a Cauchy horizon \cite{rodrigues4}. Many solutions of regular black holes have arisen since then \cite{Beato2,Beato3,Beato4,Kirill1,Kirill12,Irina,Leonardo2,balart,Nami,Ponce,Wang,silva1,Hayward,Culetu1,Culetu2,Fernando2,bambi,neves,toshmatov,azreg,DYM,ramon}, and several studies of their properties have been conduct, such as absorption \cite{macedo1,Hai1,Hai2}, scattering \cite{macedo2,Ciprian}, quasinormal modes \cite{toshmatov2,toshmatov3,lemos,li1,toshmatov5,liu2,Xi,macedo3,li2,toshmatov4,Saleh1,Wu,Pradhan,lopez}, thermodynamics \cite{Yun,Hassan,Ma,Saleh,Maluf,Wei,pacheco1,Javed,Ceren,kamiko,Abdul} and even tidal forces \cite{sharif}.

Beyond general relativity, we have the alternatives theories of gravity \cite{cap}. The Einstein equations could be obtained from the variational principle if we consider the Einstein-Hilbert action \cite{las}. Modify this action is a way to find the field equations of these alternative theories \cite{cap2}. One of the most studied modifications the $f(R)$ theory \cite{cap}, with $R$ being the curvature scalar, which emerged from Starobinsky's work in 1980, where he inserted an $R^2$ term into the Einstein-Hilbert action \cite{sta,cem}. However, we can construct other theories if we consider different curvature invariants in the action. A possible combination of these scalars is the Gauss-Bonnet term $G$ that is a topological invariant in four dimensions \cite{Nojiri1}. Although $G$ does not make modifications in the field equations, a nonlinear term of $G$ will not be a topological invariant anymore \cite{felice,rodrigues5,rodrigues6,sergei2,shamir,sergei3,bamba}. This is what we call $f(G)$ theory, which arises as an alternative to dark matter \cite{Nojiri1}.

As in general relativity, it's is possible to find regular solutions in alternative theories \cite{Lobo}. Rodrigues and collaborators generalized solutions of regular black holes from general relativity to some modified theories of gravity \cite{rodrigues1,Manuel1,Manuel2,Manuel3}. Ghosh \textit{et al.} found a solution to five dimensions when they included a linear term of $G$ in the action \cite{ghosh}.

As we said, due to the presence of charge, rotation, or a cosmological constant, new structures arise in the black holes \cite{Zerbini2}. So, in the literature, it is possible to find solutions with multiples horizons \cite{Stefano2,kirill3,Wang2,Wang3,Josef,Yun2}. An example of that is the Reissner-Nordström-de Sitter metric since we have an event, Cauchy and cosmological horizon \cite{Cardoso1}. Actually, if we consider vector-tensor theories, the number of horizons should be much greater than only three. For this type of theory, it's also possible to find metrics with multiple singularities \cite{Gao}. However, we still have the possibility of constructing regular multi-horizon black holes in general relativity and beyond if we relax some energy conditions and assume that the gravitational theory is coupled with a nonlinear electrodynamics \cite{Odintsov}.

The structure of this article is organized as follows. In the SEC. \ref{sec2} we construct the formalism for the general relativity theory, considering a spherically symmetric and static source, and we obtain the electromagnetic quantities and the energy conditions associated with the solution. In SEC. \ref{sec3} we use the same formalism and generalize the solutions to $f(R)$ gravity, finding corrections in the electromagnetic sector. In SEC. \ref{sec4} we construct a regular model, where we generalize the solution with two horizons already known from general relativity to $f(G)$ gravity in four dimensions. In SEC. \ref{sec5} we present our conclusions and discussion. The analytical expressions for $L(P)$ are obtained in Appendix \ref{ap1}. As the expressions for the anti-de Sitter example and the solution with three horizons are too complicated, the appendices \ref{ap2} and \ref{ap3} are dedicated to find the $f(R)$ and $f(G)$ functions that are generated.

\section{The formalism with general relativity}\label{sec2}
In this section, we will develop the formalism to study the solutions of regular black holes with multihorizons in the context of general relativity, so that we can generalize these results later to the alternative theories of gravity. This procedure was developed in \cite{Zerbini} and used to study multihorizon solutions in \cite{Odintsov}. The action that describes the nonlinear electromagnetic theory coupled with general relativity is
\begin{equation}
S=\int d^4x\sqrt{-g}\left[R-2\kappa^2L(I)\right],
\label{EHA}
\end{equation}
where $R$ is the curvature scalar, $g$ is  the determinant of the metric, $L$ is the electromagnetic Lagrangian and $I$ is the electromagnetic scalar, defined as $I=\frac{1}{4}F^{\mu\nu}F_{\mu\nu}$, with $F_{\mu\nu}=\partial_\mu A_\nu-\partial_\nu A_\mu$ being the Maxwell-Faraday tensor. We will consider a spherically symmetric and static spacetime described by the line element
\begin{equation}
ds^2=-e^{a(r)}dt^2+e^{-a(r)}dr^2+r^2\left(d\theta^2+\sin^2\theta d\phi^2\right).
\label{ele}
\end{equation}
To obtain the electromagnetic and gravitational field equations, we need to vary the Lagrangian \eqref{EHA} with respect to the gauge potential $A_\mu$ and the metric $g_{\mu\nu}$. These equations are 
\begin{eqnarray}
R^{\mu}_{\ \nu}-\frac{1}{2}\delta^{\mu}_{\nu}R&&=\kappa^2\left[L\delta^{\mu}_{\nu}-F^{\mu\beta}F_{\beta\nu}\partial_IL\right],\label{meq1}\\
\nabla_{\mu}\left(F^{\mu\nu}\partial_IL\right)&&=\partial_\mu\left(\sqrt{-g}F^{\mu\nu}\partial_IL\right)=0\label{meq2}\,.
\end{eqnarray}
If we consider that the source has only electric charge, the only nonzero component of the Maxwell-Faraday tensor is $F^{10}$. With this, the Eq. \eqref{meq2} becomes
\begin{eqnarray}
\partial_r\left(r^2\partial_ILF^{01}\right)=0.
\end{eqnarray}
Integrating this equation, we obtain
\begin{equation}
F^{01}=\frac{q}{r^2}\left(\partial_IL\right)^{-1},
\label{ef}
\end{equation}
where $q$ is an integration constant that represents the electric charge of the source. Using $I=-\frac{1}{2}(F^{01})^2$ and defining a new variable $X=q\sqrt{-2I}$, we get
\begin{equation}
1=r^2\partial_XL.
\end{equation}
As we can write \eqref{meq1} in the form $G^{\mu}_{\ \nu}=-\kappa^2T^{\mu}_{\;\;\nu}$ and identifying $\rho=T^{0}_{\ 0}$, $p_r=-T^{1}_{\ 1}$ and $p_t=-T^{2}_{\ 2}=-T^{3}_{\ 3}$, where $\rho$ is the energy density and $p_r$ and $p_t$ are the  radial and tangential pressures respectively, we find the following equations
\begin{eqnarray}
p_r=-\rho=L-\frac{X}{r^2},\ p_t=L\label{eq}.
\end{eqnarray}
Furthermore, from the components of the Einstein tensor, we also have the following relations
\begin{eqnarray}
re^{a} a'+e^{a}-1&=&-\kappa^2 r^2\rho\label{rho}\,,\\
re^{a} a'+e^{a}-1&=&\kappa^2 r^2p_r\label{pr}\,,\\
re^{a} a''+r e^{a} a'^2+2e^{a} a'&=&2\kappa^2 rp_t\label{pt}\,.
\end{eqnarray}
From these relations and \eqref{eq}, we have
\begin{eqnarray}
\rho=-\frac{1}{\kappa^2 r^2}\frac{d}{dr}\left[r(e^a-1)\right]\,,\,L=\frac{X}{r^2}-\rho\,,\,p_t=-\rho-\frac{r}{2}\rho'\,,\,X=-\frac{ r^3}{2}\rho'\label{eq2}\,.
\end{eqnarray}
Since we have the energy density, in terms of the radial coordinate, it is possible to find an analytical expression for $I(r)$ and invert this function to obtain $r(I)$ and consequently $L(I)$. However, as we are studying electric sources, it is not possible to find the Lagrangian. Actually, in the electric case, is more convenient to work with the auxiliary field $P_{\mu\nu}=\left(\partial_I L\right) F_{\mu\nu}$. In the Appendix \ref{ap1} we will find analytical expressions for $L(P)$ for the solutions presented here.

It is also important to verify the energy conditions that are given by the following:
\begin{enumerate}
	\item Null Energy Condition (NEC): $\rho+p_r\geq 0$ and $\rho+p_t\geq 0$;
	
	\item Weak Energy Condition (WEC): $\rho\geq 0,\rho+p_r\geq 0$ and $\rho+p_t\geq0$;
	
	\item Strong Energy Condition (SEC): $\rho+p_r+2p_t\geq 0$, $\rho+p_r\geq 0$ and $\rho+p_t\geq 0$;
	
	\item Dominant Energy Condition (DEC): $\rho\geq 0$, $\rho\pm p_r\geq 0$ and $\rho\pm p_t\geq 0$.  
\end{enumerate} 
Here we call attention to a mistake in (21) of \cite{Odintsov}, where $p_r$ is defined. By the condition $p_r=-\rho$ we expect that $p_r=L-\frac{X}{r^2}$; however in (21) of \cite{Odintsov} we see $p_r=L+\frac{X}{r^2}$. This modification changes the energy conditions, so that they also must be corrected.

In \cite{Odintsov}, three solutions were analyzed. Here we will review these solutions with the necessary corrections.

\subsection{Example in the Minkowski background}\label{subsec1}
The first model is a solution that behaves asymptotically as Minkowski and is described by
\begin{eqnarray}
e^{a(r)}=1-\frac{\alpha r^2}{\beta+r^3}=\frac{\left(r+r_0\right)\left(r-r_1\right)\left(r-r_2\right)}{\beta+r^3}\label{sol1},
\end{eqnarray} 
with
\begin{equation}
\alpha=\frac{r_1^2+r_2^2+r_1r_2}{r_1+r_2},\ \beta=\frac{r_1^2r_2^2}{r_1+r_2},\ r_0=\frac{r_1r_2}{r_1+r_2}.
\end{equation}
If we take $\alpha=2m$ and $\beta=2l^2m$, we will recover the regular black hole of Sean A. Hayward (see eq. (5) in \cite{Hayward}).

To analyze the regularity of the solution it is enough to calculate the Kretschmann scalar, $K=R^{\mu\nu\alpha\beta}R_{\mu\nu\alpha\beta}$, however, for future reasons, we will also calculate the curvature scalar and the Gauss-Bonnet invariant, $G=R^2-4R^{\mu\nu}R_{\mu\nu}+R^{\mu\nu\alpha\beta}R_{\mu\nu\alpha\beta}$,
\begin{eqnarray}
R(r)&=&\frac{6 \alpha  \beta  \left(2 \beta -r^3\right)}{\left(\beta +r^3\right)^3},\label{curhay}\\
K(r)&=&\frac{12 \alpha ^2 \left(2 \beta ^4+r^{12}-4 \beta  r^9+18 \beta ^2 r^6-2 \beta ^3 r^3\right)}{\left(\beta +r^3\right)^6},\label{krehay}\\
G(r)&=&\frac{12 \alpha ^2 \left(2 \beta ^2+r^6-6 \beta  r^3\right)}{\left(\beta +r^3\right)^4}\label{gbhay}.
\end{eqnarray}
We can see that the solution does not present divergences. From \eqref{eq2} we get
\begin{eqnarray}
\rho=-p_r=\frac{3\alpha\beta}{\kappa^2(r^3+\beta)^2}\,,\,p_t=\frac{3\alpha\beta (2r^3-\beta)}{\kappa^2(r^3+\beta)^3}\,,\,X=\frac{9\alpha\beta r^5}{\kappa^2(r^3+\beta)^3}\,, L=\frac{3\alpha\beta\left(2r^3-\beta\right)}{\kappa^2\left(r^3+\beta\right)^3}.\label{eq3}
\end{eqnarray}
\begin{figure}
	\includegraphics[height=5.cm,width=7cm]{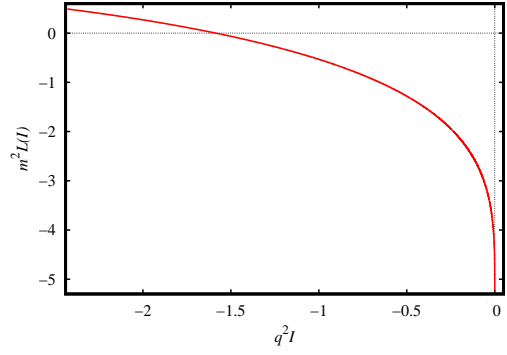}
	\caption{Electromagnetic Lagrangian for the asymptotically flat solution with $\beta=(4/3) m^3$ and $\alpha=(7/3)m$.}
	\label{LI1}
\end{figure}

The nonlinearity of the electromagnetic theory is described by the graphical representation of $L(I)$ in Fig. \ref{LI1}. 
Since we have $\rho$, $p_r$ and $p_t$, we obtain the energy conditions
\begin{eqnarray}
&&\rho\geq 0,\ \rho+p_r=0,\ \rho+p_t=\frac{9\alpha\beta r^3}{\kappa^2(r^3+\beta)^3}\geq 0,\ \rho+p_r+2p_t=\frac{6\alpha\beta (2r^3-\beta)}{\kappa^2(r^3+\beta)^3},\\
&&\rho-p_r=\frac{6\alpha\beta}{\kappa^2(r^3+\beta)^2}\geq 0,\ \rho-p_t=\frac{3\alpha\beta(2\beta-r^3)}{\kappa^2(r^3+\beta)^3}.\label{ech}
\end{eqnarray}
The weak and null energy conditions are always satisfied, however, inside the black hole the strong energy condition is violated, which is normal for regular solutions, and outside the dominant energy condition is violated, which already happened with the Bardeen solution. Now, we highlight that these results present corrections in relation to \cite{Odintsov}. From \eqref{ech}, we see that the dominant energy condition is violated for $r>(2\beta)^{1/3}$ and not $r<(2\beta)^{1/3}$ as in (26) of \cite{Odintsov}. In \cite{Odintsov}, we also see that for $r>(\beta/2)^{1/3}$, the relation $\rho-p_r$ is negative, which also violates the dominant energy condition. However, as we corrected the definition of $p_r$, that problem vanish.
\subsection{Example in the anti-de Sitter background}\label{subsec2}
\begin{figure}
		\includegraphics[height=5.cm,width=7cm]{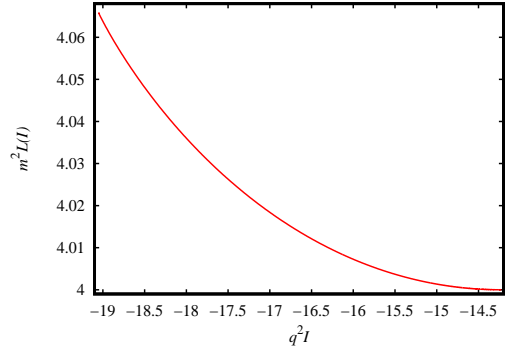}
		\caption{Graphical representation of $L(I)\times I$ associated with the asymptotically anti-de Sitter solution with $r_1=0.5m$ and $r_2=1.5m$.}
		\label{LI2}
\end{figure}

Now we will analyze the model that is asymptotically anti-de Sitter and is described by
\begin{eqnarray}
e^{a(r)}=\frac{(r^2-r_1^2)(r^2-r_2^2)}{r_1r_2(r^2+r_1r_2)}\label{sol2}\,.
\end{eqnarray}
The curvature invariants are
\begin{eqnarray}
R(r)&=&\frac{2}{r_1 r_2 \left(r^2+r_1 r_2\right)^3} \left(r^4 \left(r_1^2-16 r_1 r_2+r_2^2\right)+3 r^2 r_1 r_2 \left(r_1^2-4 r_1 r_2+r_2^2\right)+6r_1^2 r_2^2 \left(r_1^2+r_1 r_2+r_2^2\right)-6 r^6\right),\label{cs2}\\
K(r)&=&\frac{4 }{r_1^2 r_2^2 \left(r^2+r_1 r_2\right)^6}\left(6 r^{12}-2 r^{10} \left(r_1^2-16 r_1 r_2+r_2^2\right)+r^8 \left(r_1^4-8 r_1^3 r_2+72 r_1^2 r_2^2 -8r_1r_2^3 +r_2^4\right) \right. \nonumber\\
&+&4 r^6 r_1 r_2 \left(r_1^4-5 r_1^3 r_2+18 r_1^2 r_2^2-5 r_1 r_2^3+r_2^4\right)+r^4r_1^2 r_2^2 \left(19 r_1^4+20 r_1^3 r_2+92 r_1^2 r_2^2+20 r_1 r_2^3+19 r_2^4\right)\nonumber\\
&+&\left.6 r^2 r_1^3 r_2^3\left(r_1^4-3 r_1^3 r_2-2 r_1^2 r_2^2-3 r_1 r_2^3+r_2^4\right)+6 r_1^4 r_2^4 \left(r_1^2+r_1r_2+r_2^2\right)^2\right),\\
G(r)&=&\frac{8}{r_1^2 r_2^2 \left(r^2+r_1 r_2\right)^4} \left(3 r^8-r^6 \left(r_1^2-10 r_1 r_2+r_2^2\right)-2 r^4 r_1 r_2 \left(2 r_1^2-5 r_1 r_2+2 r_2^2\right)\right.\nonumber\\
&-&\left.3 r^2r_1 r_2 \left(r_1^4+7 r_1^3 r_2+8 r_1^2 r_2^2+7 r_1 r_2^3+r_2^4\right)+3 r_1^2 r_2^2\left(r_1^2+r_1 r_2+r_2^2\right)^2\right),\label{G2}
\end{eqnarray}
so that, the solution is regular in all spacetime. As in the previous examples, from \eqref{eq2} we get 
\begin{eqnarray}
\rho(r)&=&\frac{-3 r^4+r^2 \left(r_1^2-4 r_1 r_2+r_2^2\right)+3 r_1 r_2 \left(r_1^2+r_1 r_2+r_2^2\right)}{\kappa^2r_1 r_2\left(r^2+r_1 r_2\right)^2}=-p_r(r),\\
X(r)&=&\frac{r^4 (r_1+r_2)^2 \left(r^2+5 r_1 r_2\right)}{\kappa^2r_1 r_2 \left(r^2+r_1 r_2\right)^3},\\
L(r)&=&\frac{3 r^6+9 r^4 r_1 r_2+r^2 r_1 r_2 \left(r_1^2+11 r_1 r_2+r_2^2\right)-3 r_1^2 r_2^2 \left(r_1^2+r_1r_2+r_2^2\right)}{\kappa^2r_1 r_2 \left(r^2+r_1 r_2\right)^3}=p_t(r).
\end{eqnarray}
We can see that as $p_r\neq p_t$ we have the behavior of an anisotropic solution with an equation of state $\rho=-p_r$.
We highlight that the expression for $X(r)$ is different from Eqs. (29) in \cite{Odintsov}. In the infinity of the radial coordinate $L(r)$ and $X(r)$ tend to a constant and the nonlinear dependence of $L(I)$ is represented in \ref{LI2}. The energy conditions are
\begin{eqnarray}
&&\rho=\frac{-3 r^4+r^2 \left(r_1^2-4 r_1 r_2+r_2^2\right)+3 r_1 r_2 \left(r_1^2+r_1 r_2+r_2^2\right)}{\kappa^2r_1 r_2\left(r^2+r_1 r_2\right)^2}, \rho+p_r=0, \rho+p_t=\frac{r^2 (r_1+r_2)^2 \left(r^2+5 r_1 r_2\right)}{\kappa^2r_1 r_2 \left(r^2+r_1 r_2\right)^3},\nonumber\\ 
&&\rho+p_r+2p_t=\frac{2 \left(3 r^6+9 r^4 r_1 r_2+r^2 r_1 r_2 \left(r_1^2+11 r_1 r_2+r_2^2\right)-3 r_1^2 r_2^2
	\left(r_1^2+r_1 r_2+r_2^2\right)\right)}{\kappa^2r_1 r_2 \left(r^2+r_1 r_2\right)^3},\label{ecad}\\
&&\rho-p_r=\frac{-6 r^4+2 r^2 \left(r_1^2-4 r_1 r_2+r_2^2\right)+6 r_1 r_2 \left(r_1^2+r_1 r_2+r_2^2\right)}{\kappa^2r_1 r_2
	\left(r^2+r_1 r_2\right)^2},\nonumber\\
&&\rho-p_t=\frac{-6 r^6+r^4 \left(r_1^2-16 r_1 r_2+r_2^2\right)+3 r^2 r_1 r_2 \left(r_1^2-4 r_1 r_2+r_2^2\right)+6 r_1^2
	r_2^2 \left(r_1^2+r_1 r_2+r_2^2\right)}{\kappa^2r_1 r_2 \left(r^2+r_1 r_2\right)^3}.\nonumber
\end{eqnarray}
We can see that the energy density is not always positive, which is not a problem since we are considering a model that is asymptotically anti-de Sitter so that, we have the contribution of the negative cosmological constant, which can be regarded as a negative energy density. For points outside the event horizon, WEC and DEC are violated while SEC is violated inside the event horizon.
\subsection{Regular black hole with three horizons}\label{subsec3}
\begin{figure}
		\includegraphics[height=5.cm,width=7cm]{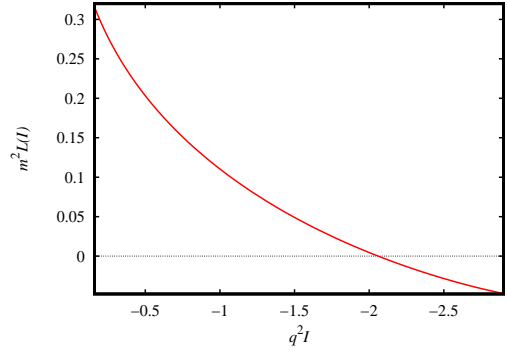}
		\caption{Behavior of the electromagnetic Lagrangian in terms of $I$ associated to the solution with three horizon, for $r_1=0.5m$, $r_2=1.5m$, $r_3=16m$ and $l=20m$.}
		\label{LI3}
\end{figure}

Now we will use the formalism that we developed in the previous section to study the multihorizon solution. We consider the configuration described by
\begin{equation}
e^{a(r)}=-\frac{(r^2-r_1^2)(r^2-r_2^2)(r^2-r_3^2)}{l^2\left(r^2+\frac{r_1r_2r_3}{l}\right)^2}.\label{sol3}
\end{equation}
If we expand this solution near the origin and in infinity of the radial coordinate we get
\begin{eqnarray}
e^{a(r)} \approx 1-r^2 \left(\frac{2 l}{\gamma}+\frac{\delta}{\gamma^2}\right)+O\left(r^3\right),\ \mbox{for}\ r\rightarrow 0,\\
e^{a(r)} \approx -\frac{r^2}{l^2}+\frac{l \omega+2 \gamma}{l^3}+O\left(\frac{1}{r^2}\right),\ \mbox{for}\ r\rightarrow \infty,
\end{eqnarray}
with
\begin{equation}
\omega=r_1^2 +r_2^2+r_3^2, \ \gamma=r_1r_2r_3, \ \delta=r_1^2r_2^2+r_1^2r_3^2+r_2^2r_3^2.
\end{equation}
The solution behaves like de Sitter for points inside and far from the event horizon ($r_2$). The curvature and topological invariants are
\begin{eqnarray}
R(r)&=&\frac{2}{\left(l r^2+\gamma\right)^4} \left(l^4 r^6+4 l^3 r^4 \gamma+l^2 \left(6 r^8-r^6 \omega+3 \gamma^2\right)+2 l \gamma \left(11 r^6-2 r^4 \omega-3 r^2 \delta+6 \gamma^2\right)+\gamma^2 \left(28 r^4-15 r^2\omega+6 \delta\right)\right),\nonumber\\ \\
K(r)&=&\frac{4}{\left(l r^2+\gamma\right)^8} \left(l^8 r^{12}+8 l^7 \gamma r^{10}+2 l^6 \left(r^6-\omega r^4+\delta r^2+13 \gamma^2\right) r^8\right.+4 l^5 \gamma \left(3 r^6-3\omega r^4+3 \delta r^2+11 \gamma^2\right) r^6\nonumber\\
&+&l^4 \left(6 r^{12}-2 \omega r^{10}+\omega^2 r^8-2\left(\delta\omega-12\gamma^2\right) r^6+14 \left(\delta^2-2\omega\gamma^2\right) r^4-48 \gamma^2\delta r^2+157 \gamma^4\right) r^4+4 l^3\gamma \left(11 r^{12}\right.\nonumber\\
&-&\left.3 \omega r^{10}+\left(\omega^2-4\delta\right) r^8+\left(11 \delta\omega+9\gamma^2\right) r^6-\left(11\delta^2+46\omega\gamma^2\right) r^4+47 \gamma^2\delta r^2+3 \gamma^4\right) r^2+\gamma^4 \left(262 r^8-230 \omega r^6\right.\nonumber\\
&+&\left.\left(53 \omega^2+56 \delta\right) r^4-30 \delta\omega r^2+6 \delta^2\right)+4 l \gamma^3 \left(55 r^{10}+3 \omega r^8-3 \left(3\omega^2+16 \delta\right) r^6+7 \left(3 \delta\omega\right.+\gamma^2\right) r^4-3\left(\delta^2+5 \omega\gamma^2\right) r^2\nonumber\\
&+&\left.6 \gamma^2\delta\right)+2 l^2 \gamma^2 \left(70 r^{12}-28 \omega r^{10}+\left(29 \omega^2+12 \delta\right.\right) r^8-\left(64 \delta\omega+141\gamma^2\right) r^6+\left(34 \delta^2+31\gamma^2\omega\right) r^4-9 \gamma^2\delta r^2+\left.\left.12 \gamma^4\right)\right),\nonumber\\ \\
G(r)&=&\frac{8}{\left(l r^2+\gamma\right)^6} \left(l^4 r^4 \left(r^6+3 r^2 \delta-10 \gamma^2\right)+2 l^3 r^2\gamma \left(3 r^6+3 r^4 \omega-r^2 \delta-9 \gamma^2\right)+l^2 \left(3 r^{12}-r^{10} \omega-3 r^6 \left(\delta\omega-7\gamma^2\right)\right.\right.\nonumber\\
&+&\left.r^4 \left(5\delta^2+16\omega\gamma^2\right)-33 r^2\gamma^2\delta+12 \gamma^4\right)+2 l \gamma \left(8 r^{10}-3 r^8
	\omega-3 r^6 \left(\omega^2+2\delta\right)+r^4 \left(11\delta\omega+28\gamma^2\right)\right.\nonumber\\
&-&\left.\left.3 r^2 \left(2\delta^2+5\omega\gamma^2\right)+6 \gamma^2\delta\right)+\gamma^2 \left(33 r^8-45 r^6\omega+14 r^4 \left(\omega^2+2\delta \right)-15 r^2 \delta\omega+3\delta^2\right)\right).\label{G3}
\end{eqnarray} 
These functions are complicated, however, it is not difficult to see the regularity at the origin. In infinity of the radial coordinate, we get
\begin{eqnarray}
\lim\limits_{r\rightarrow \infty}\left\{R(r),K(r),G(r)\right\}=\left\{\frac{12}{l^2},\frac{24}{l^4},\frac{24}{l^4}\right\},
\end{eqnarray} 
which are viable results since the solution is asymptotically de Sitter. The fluid quantities, $X(r)$ and the electromagnetic Lagrangian are
\begin{eqnarray}
\rho(r)&=&\frac{1}{\kappa^2\left(l r^2+\gamma\right)^{3}}\left[l^3 r^4+3 l^2 r^2 \gamma+l \left(3 r^6-r^4\omega-r^2 \delta+6 \gamma^2\right)+\gamma \left(7 r^4-5 r^2 \omega+3 \delta\right)\right]=-p_r(r),\label{rho3}\\
X(r)&=&\frac{r^4}{\kappa^2\left(l r^2+\gamma\right)^{4}}\left[l^4 r^4+4 l^3 r^2 \gamma-l^2 \left(r^4 \omega+\right.2 r^2 \delta-15 \gamma^2\right)-2 l \gamma \left.\left(r^4+4 r^2\omega-5 \delta\right)+\gamma^2 \left(5\omega-14 r^2\right)\right],\label{X3}\\
L(r)&=&\frac{1}{\kappa^2\left(l r^2+\gamma\right)^{4}}\left[l^2 \left(-\left(3 r^8+r^4 \delta-6 r^2 \gamma^2\right)\right)\right.-2 l
\gamma \left(6 r^6+r^4 \omega-4 r^2 \delta\left.+3 \gamma^2\right)+\gamma^2 \left(-21 r^4+10 r^2 \omega-3\delta\right)\right]=p_t(r).\nonumber\\
\end{eqnarray}
We see that $\rho$, in (44) of \cite{Odintsov} and $X$ in (45) of \cite{Odintsov}, differ from what we obtained here, \eqref{rho3} and \eqref{X3}, so that is expected that the Lagrangian $L_{BCX}$ is different from (46) in \cite{Odintsov}. The nonlinearity of the electromagnetic theory that generates this metric is described in Fig. \ref{LI3}. As in the solution before, $L(r)$ and $X(r)$ tend to a constant to $r\rightarrow \infty$.
Together with \eqref{rho3}, the energy conditions are
\begin{eqnarray}
&&\rho+p_r=0,\\
&&\rho+p_t=\frac{r^2}{\kappa^2 \left(l r^2+\gamma\right)^{4}}\left[l^4 r^4+4 l^3 r^2 \gamma-l^2 \left(r^4 \omega+2 r^2 \delta-15 \gamma^2\right)
-2 l \gamma \left(r^4+4 r^2\omega-5 \delta\right)+\gamma^2 \left(5\omega-14 r^2\right)\right],\\
&&\rho+p_r+2p_t=-\frac{2}{\kappa^2\left(l r^2+\gamma\right)^{4}}\left[l^2 \left(3 r^8+r^4 \delta-6 r^2 \gamma^2\right)+2 l
\gamma \left(6 r^6+r^4 \omega-4 r^2 \delta+3 \gamma^2\right)-\gamma^2 \left(21 r^4-10 r^2\omega+3\delta\right)\right],\nonumber\\\\
&&\rho-p_r=\frac{2}{\kappa^2 \left(l r^2+\gamma\right)^{3}}\left[l^3 r^4+3 l^2 r^2 \gamma+l \left(3 r^6-r^4\omega-r^2 \delta+6 \gamma^2\right)+\gamma \left(7 r^4-5 r^2 \omega+3 \delta\right)\right],\\
&&\rho-p_t=\frac{1}{\kappa^2\left(l r^2+\gamma\right)^{4}}\left[l^4 r^6+4 l^3 r^4 \gamma+l^2 \left(6 r^8-r^6 \omega+3 r^2 \gamma^2\right)+2 l
\gamma \left(11 r^6-2 r^4 \omega-3 r^2 \delta+6 \gamma^2\right)\right.\nonumber\\
&&+\left.\gamma^2 \left(28 r^4-15 r^2 \omega+6\delta\right)\right].
\end{eqnarray}
We can see that there is the possibility for all energy conditions to be violated.
\section{Regular multihorizon black Holes in $f(R)$ gravity with nonlinear Electrodynamics}\label{sec3}
Now we will consider a line element written as
\begin{equation}
ds^2=-e^{a(r)}dt^2+e^{b(r)}dr^2+r^2\left(d\theta^2+\sin^2\theta d\phi^2\right).
\label{ele2}
\end{equation}
When we have nonlinear electrodynamics coupled with $f(R)$ gravity, the action is given by
\begin{equation}
S_{f(R)}=\int d^4x\sqrt{-g}\left[f(R)-2\kappa^2L(I)\right].
\end{equation}
If we vary this action with respect to the metric we get the equations of motion for $f(R)$ gravity, which can be written as
\begin{eqnarray}
&&\hspace{-.3cm}R_{\mu\nu}-\frac{1}{2}g_{\mu\nu}R=-
f_R^{-1}\big[\kappa^2 T_ { \mu\nu }
-\frac{1}{2}g_{\mu\nu}\left(f-Rf_R\right) -\left(g_{\mu\nu}\square-\nabla_{
	\mu}\nabla_{\nu}\right)f_R\big] \equiv -\kappa^ { 2
}  \mathcal{T}_{\mu\nu}^{(eff)}\label{energyeff}.
\end{eqnarray}
Identifying $\mathcal{T}_{0}^{0(eff)}=\rho^{(eff)}$, $\mathcal{T}_{1}^{1(eff)}=-p_{r}^{(eff)}$, and $\mathcal{T}_{2}^{2(eff)}=\mathcal{T}_{3}^{3(eff)}=-p_{t}^{(eff)}$, we can write the energy conditions for $f(R)$ gravity as \cite{Reboucas,Cembranos1,Cembranos2,Capozziello}
\begin{eqnarray}
&&NEC_{1,2}(r)=\rho^{(eff)}+p_{r,t}^{(eff)}\geq 0\;,\label{cond1}\\
&&SEC(r)=\rho^{(eff)}+p_{r}^{(eff)}+2p_{t}^{(eff)}\geq 0\,,\label{sec}\\
&&WEC_{1,2}(r)=\rho^{(eff)}+p_{r,t}^{(eff)}\geq 0\;,\label{cond2}\\
&&DEC_{1}(r)=\rho^{(eff)}\geq 0, \label{cond3a}\\
&&DEC_{2,3}(r)=\rho^{(eff)}-p_{r,t}^{(eff)}\geq 
0\;,\label{cond3}
\end{eqnarray}
where, for nonlinear electrodynamics, we have
\begin{eqnarray}
&&\rho^{(eff)}=-\frac{e^{-b}}{4\kappa^2r^2f_R}\Bigg\{4\kappa^2r^2\left[F^{10 } \right]^2e^{a+2b}\partial_IL+4\kappa^2r^2e^bL 
+4r^2\frac{d^2f_R}{dr^2} +\left(8r-2r^2b'\right)\frac{df_R}{dr}\nonumber\\
&& \qquad\quad +\big[(r^2a'+4r)b'+4e^b-2r^2a''-r^2\left(a'\right)^2 -4ra'-4\big] f_R +2r^2e^bf \Bigg\} \label{rhoeff}\;,\\
&& p_{r}^{(eff)}= \frac{e^{-b}}{4\kappa^2r^2f_R}\Bigg\{4\kappa^2r^2\left[F^{10}\right]^2e^{a+2b}\partial_IL+4\kappa^2r^2e^bL+(2r^2a'+8r)\frac{df_R}{dr} +\big[ (r^2a'+4r)b'\nonumber\\
&&\qquad\quad +4e^b-2r^2a''-r^2\left(a'\right)^2-4ra'-4\big]f_R+2r^2e^bf\Bigg\}\;,\label{preff}\\
&& p_{t}^{(eff)}=\frac{e^{-b}}{4\kappa^2r^2f_R}\Bigg\{4\kappa^2r^2e^bL+4r^2\frac{d^2f_R}{dr^2}+[2r^2(a'-b')+4r]\frac{df_R}{dr}+\big[(r^2a'+4r)b'\nonumber\\
&&\qquad\quad  +4e^b-2r^2a''-r^2\left(a'\right)^2-4ra'-4\big]f_R+2r^2e^bf\Bigg\}\;.\label{pteff}
\end{eqnarray}
In view of the identity $WEC_3(r)\equiv DEC_1(r)$, one of the conditions is not written. In \cite{Manuel2}, the following theorem was proven: Given a solution of Eqs. \eqref{energyeff} of $f(R)$ gravity described by $S_1=\{a(r),b(r),f(R),L,F^{10}(r)\}$, if there exists a solution in general relativity  $S_2=\{a(r),b(r),\bar{L},\bar{F}^{10}(r)\}$, then the energy conditions \eqref{cond1}-\eqref{cond3} are identical for $S_1$ and $S_2$. This implies that by, taking the models \eqref{sol1}, \eqref{sol2} and \eqref{sol3} for the $f(R)$ gravity, the energy conditions \eqref{cond1}-\eqref{cond3} are the same that we have in the general relativity cases.

Even if there are no changes in the energy conditions, due to the coupling with $f(R)$ gravity, we have modifications on the structure of $X(r)$ and $L$. The components of the field equations for $f(R)$ gravity, considering $a(r)=-b(r)$, are
\begin{eqnarray}
\frac{e^{a} \left(\left(-r a'-4\right) f_R'+f_R \left(r a''+r a'^2+2 a'\right)-2 r f_R''\right)}{r}-f(R)-2 \kappa ^2\rho=0,\label{eqfR1}\\
\frac{e^{a} \left(\left(-r a'-4\right) f_R'+f_R \left(r a''+r a'^2+2 a'\right)\right)}{r}-f(R)+2\kappa ^2 p_r=0,\label{eqfR2}\\
-\frac{e^{a} \left(\left(r a'+1\right) f_R'+r f_R''\right)}{r}+\frac{f_R \left(r e^{a} a'+e^{a}-1\right)}{r^2}-\frac{f(R)}{2}+\kappa ^2p_t=0.\label{eqfR3}
\end{eqnarray}
In the field equations, we need to pay attention that we are using $\rho$, $p_r$ and $p_t$ and not $\rho^{(eff)}$, $p_r^{(eff)}$ and $p_t^{(eff)}$ as in the energy conditions.

Subtracting \eqref{eqfR1} from \eqref{eqfR2} we get
\begin{equation}
e^a f''_R+\kappa^2\left(\rho+p_r\right)=0.
\end{equation}
To obtain $\rho=-p_r$, with $e^a\neq 0$, we need $f''_R=0$, so that, $f_R$ is
\begin{equation}
f_R=c_0+c_1 r,
\end{equation}
where $c_0$ and $c_1$ are integration constants. Since we have $R(r)$, we can invert this function and obtain $r(R)$ and then we find $f_R(R)$. The $f(R)$ function is 
\begin{equation}
f(R)=c_0R+c_1 \int r(R)dR,
\end{equation}
or, in terms of the radial coordinate,
\begin{equation}
f(R)=\int f_R \frac{dR}{dr}dr.
\end{equation}

Using the models \eqref{sol1}, \eqref{sol2} and \eqref{sol3}, we can find the expressions for $\rho$, $L$ and $X$ in $f(R)$ gravity.
\subsection{Example in the Minkowski background}\label{subsec4}
Considering the model \eqref{sol1} we have the curvature scalar \eqref{curhay}. Inverting this function, we get that $f_R$ is
\begin{equation}
f_R(R)=c_0+c_1\left[\frac{\alpha_1(R)}{R}+\frac{2\alpha\beta}{\alpha_1(R)}-\beta\right]^{1/3},
\end{equation}
with
\begin{equation*}
\alpha_1(R)=\left\{\alpha \beta  R\sqrt{\beta R (81 \beta  R-8 \alpha )}-9 \alpha  \beta ^2 R^2\right\}^{1/3}.
\end{equation*}
Integrating $f_R$ with respect to the curvature scalar we get
\begin{eqnarray}
f(R)&=&c_0R-\frac{c_1\alpha}{\beta^{2/3}}\Bigg\{2 \sqrt{3} \tan ^{-1}\left[\frac{1}{\sqrt{3}}\left(1-2 \sqrt[3]{ \frac{2 \alpha }{\alpha_1(R)}-1+\frac{\alpha_1(R)}{\beta R}}\right)\right]-2 \ln \left(1+\sqrt[3]{  \frac{2 \alpha }{\alpha_1(R)}-1+\frac{\alpha_1(R)}{\beta R}}\right)\nonumber\\
&+&\ln \left(1	-\sqrt[3]{ \frac{2 \alpha }{\alpha_1(R)}-1+\frac{\alpha_1(R)}{\beta R}}+\left(\frac{2
		\alpha }{\alpha_1(R)}-1+\frac{\alpha_1(R)}{\beta R}\right)^{2/3}\right)-\frac{1}{72\ 2^{2/3}R^2 \alpha ^3 \beta ^3}\times\nonumber\\
&&\sqrt[3]{\frac{2\alpha_1(R)\beta^2\left(2 \alpha -9 \alpha_1(R)\right)}{R \alpha }-4 \beta^3-\frac{\alpha_1^{5}(R)}{R^3 \alpha ^2}} \left(72 R^3 \alpha ^2 \beta ^3+R^2 \alpha  \left(32 \alpha ^2+60	\alpha_1(R) \alpha +27 \alpha_1^{2}(R)\right) \beta ^2\right.\nonumber\\
&-&\left.16 R \alpha ^2 \alpha_1^{2}(R) \beta + \alpha_1(R)(\alpha_1^3(R)+9 \alpha  \beta ^2 R^2) \left(8 \alpha +3 \alpha_1(R)\right)\right)\Bigg\}.
\end{eqnarray}
If we consider the corrections from $f(R)$ gravity $\rho(r)$, $X(r)$ and $L(r)$ are
\begin{eqnarray}
\rho(r)&=&\frac{3 \alpha  \beta  c_0}{\kappa ^2 \left(\beta +r^3\right)^2}+\frac{c_1}{2 \kappa ^2} \left(\frac{6 \alpha  r}{\beta +r^3}+\frac{\alpha}{\beta ^{2/3}}  \left(\ln \left(\beta ^{2/3}+r^2-\sqrt[3]{\beta }
	r\right)-2 \ln \left(\sqrt[3]{\beta }+r\right)+2 \sqrt{3} \tan ^{-1}\left(\frac{1-\frac{2 r}{\sqrt[3]{\beta }}}{\sqrt{3}}\right)\right)\right.\nonumber\\
&-&\left.\frac{6 \alpha  r^4}{\left(\beta
	+r^3\right)^2}-\frac{4}{r}\right),\\
L(r)&=&\frac{3 \alpha  \beta  c_0 \left(2 r^3-\beta \right)}{\kappa ^2 \left(\beta +r^3\right)^3}-\frac{c_1}{2 \kappa ^2} \left(\frac{6 \alpha  r}{\beta +r^3}+\frac{\alpha}{\beta
	^{2/3}}  \left(\ln \left(\beta	^{2/3}+r^2-\sqrt[3]{\beta } r\right)-2 \ln \left(\sqrt[3]{\beta }+r\right)+2 \sqrt{3} \tan ^{-1}\left(\frac{1-\frac{2 r}{\sqrt[3]{\beta }}}{\sqrt{3}}\right)\right)\right.\nonumber\\
&+&\left.\frac{18 \alpha  r^7}{\left(\beta +r^3\right)^3}-\frac{27 \alpha  r^4}{\left(\beta +r^3\right)^2}-\frac{2}{r}\right),\\
X(r)&=&\frac{9 \alpha  \beta  c_0 r^5}{\kappa ^2 \left(\beta +r^3\right)^3}-\frac{c_1 r \left(2 \beta ^3+r^8 (2 r-3 \alpha )+3 \beta  r^5 (2 r-7 \alpha )+6 \beta ^2 r^3\right)}{2 \kappa^2 \left(\beta +r^3\right)^3}.\label{XfR}
\end{eqnarray}
It is easy to see that if we take the limit $c_1\rightarrow 0$ and $c_0\rightarrow1$, we recover the results presented in \ref{subsec1}. From \eqref{XfR}, we can see that, different from general relativity, due to the coupling with $f(R)$ gravity, $X(r)$ diverges in the infinity.

In Einstein's theory, the energy density from the energy conditions is equal to the electromagnetic energy density. When we consider $f(R)$ gravity these two quantities are not the same. Here, in the energy conditions \eqref{cond3} we have $\rho^{(eff)}$ which is the same from \eqref{eq3}, which is always positive, while the electromagnetic energy density is given by $\rho$, which has negative contributions due to the coupling with $f(R)$ theory.

With this example, we may realize the modifications that appear in the electromagnetic quantities because of $f(R)$ gravity. It is also possible to apply this formalism to the other examples. However, as the expressions are much more complicated, we will do this in the appendix \ref{ap2}, where we obtain analytical expressions for $f_R$, $X$, $\rho$ and $L$. 

\section{Regular multihorizon solutions in $f(G)$ gravity}\label{sec4}
In this section, we will analyze the possibility of generalizing the solutions for $f(G)$ theory in four dimensions. The action that describes this theory is
\begin{equation}
S_{f(G)}=\int d^4x \sqrt{-g}\left[R+f(G)-2\kappa^2 L(I)\right].
\end{equation}
Varying this action with respect to the metric we get the field equations for $f(G)$ theory, which are written as
\begin{equation}
R_{\mu\nu}-\frac{1}{2}g_{\mu\nu}R+8\left[R_{\mu\alpha\nu\beta}+R_{\alpha\nu}g_{\beta\mu}-R_{\alpha\beta}g_{\nu\mu}-R_{\mu\nu}g_{\beta\alpha}+R_{\mu\beta}g_{\nu\alpha}\right]\nabla^\alpha\nabla^\beta f_G+\left(Gf_G-f\right)g_{\mu\nu}=-\kappa^2T_{\mu\nu},\label{fe}
\end{equation}
where the subscript $G$ denotes the derivation with respect to the Gauss-Bonnet term. To analyze the energy conditions, we rewrite the field equations as
\begin{equation}
G_{\mu\nu}=-\kappa^2T_{\mu\nu}-8\left[R_{\mu\alpha\nu\beta}-R_{\alpha\nu}g_{\beta\mu}-R_{\alpha\beta}g_{\nu\mu}-R_{\mu\nu}g_{\beta\alpha}+R_{\mu\beta}g_{\nu\alpha}\right]\nabla^\alpha\nabla^\beta f_G-\left(Gf_G-f\right)g_{\mu\nu}=-\kappa^2 \mathcal{T}^{(eff)}_{\mu\nu},
\label{Geff}
\end{equation}
where the components of $\mathcal{T}_{\mu\nu}^{(eff)}$ may be identified by
\begin{equation}
\mathcal{T}_{\mu\nu}^{(eff)}=diag\left(\rho^{(eff)}(r),-p^{(eff)}_r(r),-p^{(eff)}_t(r),-p^{(eff)}_t(r)\right),
\label{Teff}
\end{equation}
where $\rho^{(eff)}(r)$, $p^{(eff)}_r(r)$ and $p^{(eff)}_t(r)$ are the effective energy density, radial pressure and tangential pressure, respectively, with contributions from $f(G)$ theory. The energy conditions are given by \cite{bamba}
\begin{eqnarray}
SEC(r)&=&\rho^{(eff)}+p^{(eff)}_r+2p^{(eff)}_t \geq 0,\label{EC1}\\
WEC_{1,2}(r)&=&NEC_{1,2}(r)=\rho^{(eff)}+p^{(eff)}_{r,t}\geq 0,\label{EC2}\\
WEC_{3}(r)&=&DEC_{1}(r)=\rho^{(eff)} \geq 0,\label{EC3}\\
DEC_{2,3}(r)&=&\rho^{(eff)}-p^{(eff)}_{r,t} \geq0.\label{EC4}
\end{eqnarray}
As in the case of $f(R)$ theory, it was shown that due to the fact that the energy conditions are taken on the Einstein tensor, the energy conditions in $f(G)$ theory are the same as general relativity \cite{Manuel3}.

The nonzero components of the field equations, to a line element as \eqref{ele2} with $a=-b$, are
\begin{eqnarray}
\frac{e^{a} \left(4 f_G a''+a' \left(4 f_G a'-4 f_G'+r\right)-8 f_G''+1\right)-4 e^{2 a} \left(-3 a' f_G'+f_G
	\left(a''+2 a'^2\right)-2 f_G''\right) -1}{r^2}+ f(G)+\kappa ^2 \rho=0,&&\label{eqfG1}\\
\frac{e^{a} \left(4 f_G a''+a' \left(4 f_G a'-4 f_G'+r\right)+1\right)-4 e^{2 a} \left(f_G \left(a''+2 a'^2\right)-3 a'
	f_G'\right)-1}{r^2}+ f(G)-\kappa ^2 p_r=0,&&\label{eqfG2}\\
\frac{e^{a} \left(a'' \left(8 e^{a} \left(r f_G'-f_G\right)+8 f_G+r^2\right)+2 r a' \left(4 e^{a} f_G''+1\right)+a'^2 \left(8 f_G+r^2-16e^{a} \left(f_G-r f_G'\right)\right)\right)}{2 r^2}+f(G)-\kappa ^2 p_t=0.\label{eqfG3}&&
\end{eqnarray}
Subtracting \eqref{eqfG1} from \eqref{eqfG2} we obtain
\begin{equation}
\kappa^2\left(\rho+p_r\right)+\frac{8e^a\left(e^a-1\right)f_G''}{r^2}=0.
\end{equation}
As $e^a\neq 1$ and $e^a\neq 0$, in general, we need $f_G''=0$ to obtain the condition $\rho=-p_r$. Consequently, $f_G$ is
\begin{equation}
f_G=c_0+c_1 r.
\end{equation}
As we did in $f(R)$ theory, we need to find an expression for $r=r(G)$ and with that construct the functions $f_G(G)$ and $f(G)$ by
\begin{equation}
f(G)=c_0 G+c_1 \int r(G)dG,
\end{equation}
or, in terms of the radial coordinate,
\begin{equation}
f(G)=\int f_G\frac{dG}{dr}dr.
\end{equation}
\subsection{Example in the Minkowski background}\label{subsec5}
\begin{figure}
	\includegraphics[height=6.cm,width=9cm]{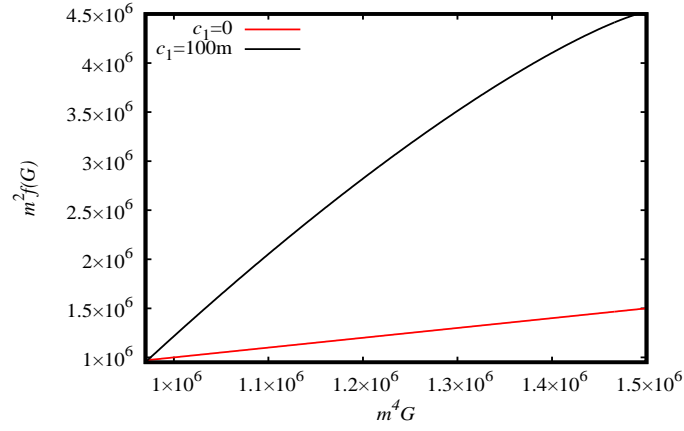}
	\caption{Difference of the $f(G)$ function to a linear and a nonlinear term of $G$ that generate the solution \eqref{sol1} with $q=0.1m$.}
	\label{fghw}
\end{figure}
In \cite{Manuel3}, the $f_G(G)$ and $f(G)$ functions were constructed for the models \eqref{sol1}, (A8) and (A10) in \cite{Manuel3}, with $\alpha=2m$ and $\beta=q^3$, where, due to astrophysical constraints, $q^2\ll m^2$ \cite{Punsly}, with $m$ being de ADM mass. We expected that the ADM mass to assume values of $10^{-17}M_{\odot}$, for primordial black holes \cite{Bellomo}, up to $10^{9}M_{\odot}$ for supermassive black holes \cite{eventhorizon}. In the Fig. \ref{fghw} we plot the $f(G)$ function that generates the solution and we compare to the linear function. Graphically these functions are different and, for some range of $G$, the intensity to $c_1=0$ is smaller. Even if there are no changes in the energy conditions, we will get corrections for the functions $\rho(r)$, $X(r)$ and $L$. These functions are given by
\begin{eqnarray}
\rho(r)&=&\frac{3 \alpha  \beta }{\kappa ^2 \left(\beta +r^3\right)^2}+\frac{\alpha  c_1}{3 \beta ^{5/3} \kappa ^2} \left(\frac{12 \beta ^{2/3}}{r \left(\beta +r^3\right)^3} \left(4 \beta ^3+\alpha  r^8+\beta  r^5 (5 \alpha -2 r)+2 \beta ^2 r^2 (r-\alpha )\right)\right.\nonumber\\
&-&\left.4	\alpha  \left(\ln \left(\beta ^{2/3}+r^2-\sqrt[3]{\beta } r\right)-2 \ln \left(\sqrt[3]{\beta }+r\right)+2 \sqrt{3} \tan ^{-1}\left(\frac{1-\frac{2 r}{\sqrt[3]{\beta
	}}}{\sqrt{3}}\right)\right)\right)=-p_r,\\
L(r)&=&\frac{3 \alpha  \beta  \left(2 r^3-\beta\right)}{\kappa ^2 \left(\beta
	+r^3\right)^3}+\frac{4 \alpha c_1 }{3 \beta ^{5/3} \kappa ^2}\left(\alpha  \left(\ln \left(\beta ^{2/3}+r^2-\sqrt[3]{\beta } r\right)-2 \ln \left(\sqrt[3]{\beta }+r\right)+2 \sqrt{3} \tan^{-1}\left(\frac{1-\frac{2 r}{\sqrt[3]{\beta }}}{\sqrt{3}}\right)\right)\right.\nonumber\\
&-&\left.\frac{3 \beta ^{2/3} \left(2 \beta ^4+\alpha  r^{11}+2 \beta  r^9-12 \beta ^2 r^5 (r-2 \alpha )-2
		\beta ^3 r^2 (\alpha +6 r)\right)}{r \left(\beta +r^3\right)^4}\right)=p_t,\\
X(r)&=&\frac{9 \alpha	\beta  r^5}{\kappa ^2 \left(\beta +r^3\right)^3}+\frac{4 \alpha  c_1 r \left(2 \beta ^3-4 r^9+6 \alpha  r^8+12 \beta  r^6-21 \alpha  \beta  r^5+18 \beta ^2 r^3\right)}{\kappa ^2 \left(\beta +r^3\right)^4}.
\end{eqnarray}

\begin{figure}
	\includegraphics[height=6.cm,width=9cm]{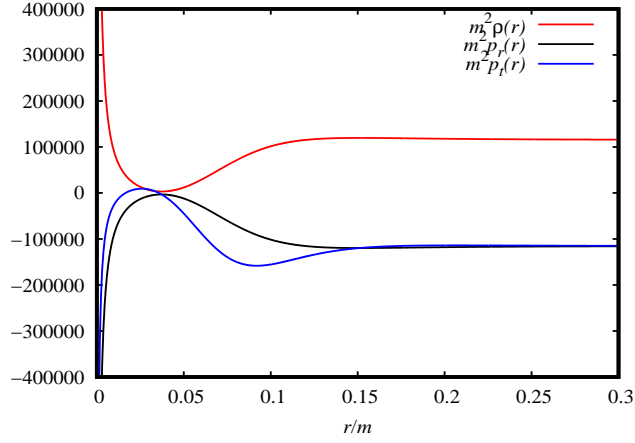}
	\caption{Behavior of the electromagnetic fluid quantities in terms of the radial coordinate with $c_1=1m$ and $q=0.1m$.}
	\label{fluid}
\end{figure}

\begin{figure}
	\includegraphics[height=6.cm,width=9cm]{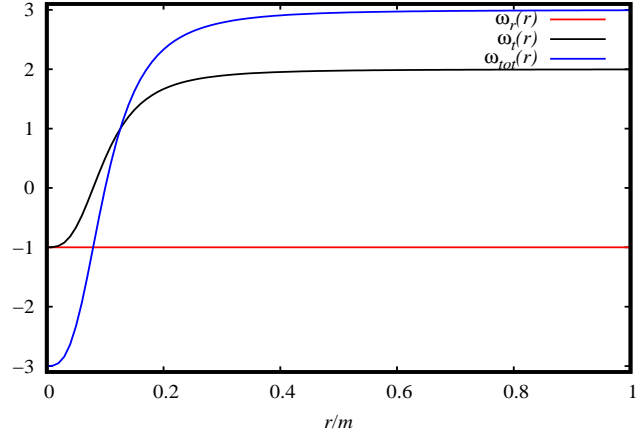}
	\caption{Parameters of the equations of state in terms of the radial coordinate with $q=0.1m$.}
	\label{ws}
\end{figure}

We realize that the results of general relativity are recovered with $c_1\rightarrow 0$ since $c_0$ does not appear. The constant $c_0$ does not appear in the equations due to the fact that it follows the linear term in the Gauss-Bonnet invariant, which in four dimensions does not make modifications in the field equations. Different from $f(R)$ gravity, $X$ is well behaved for $r\rightarrow \infty$. From SEC. \ref{sec2} we know that $I=-\frac{1}{2}(F^{10})^2$ and $X=q\sqrt{-2I}$, which implies the electric field and $X$ have the same behavior in relation to the radial coordinate, so that, $F^{10}$ is well-behaved in all spacetime and goes to zero in the black hole center and in the infinity. In Fig. \ref{fluid}, we plot the electromagnetic energy density and pressures as functions of the radial coordinate. The energy density presents a minimal value where the radial pressure has a maximum, while the tangential pressure presents minimum and maximum. In general relativity, normally, to an electrically charged regular solution, we have the behavior of an anisotropic fluid but near to the black hole center, it behaves approximately as an isotropic fluid, which is different in $f(G)$ gravity. Expanding these functions near to the center, we find
\begin{eqnarray}
\rho(r)&\approx& \frac{32 c_1 m}{\kappa ^2 q^3 r}+\frac{2 m \left(-8 \pi 
		c_1 m+9\sqrt{3} q^{2}\right)}{3\sqrt{3} \kappa ^2 q^{5}}-\frac{80 c_1 mr^2}{\kappa ^2 q^6}+O\left(r^3\right),\\
p_r(r)&\approx& -\frac{32 c_1 m}{\kappa ^2 q^3 r}+\frac{2 m \left(8 \pi 
	c_1 m-9\sqrt{3} q^{2}\right)}{3\sqrt{3} \kappa ^2 q^{5}}+\frac{80 c_1 mr^2}{\kappa ^2 q^6}+O\left(r^3\right),\\
p_t(r)&\approx&-\frac{16 c_1 m}{\kappa ^2 q^3r}+\frac{2 m \left(8 \pi  c_1 m-9\sqrt{3} q^{2}\right)}{3\sqrt{3} \kappa ^2 q^{5}}+\frac{160 c_1 m r^2}{\kappa
	^2 q^6}+O\left(r^3\right).
\end{eqnarray}
The term that arises from $f(G)$ theory is responsible for the anisotropy in the black hole center. In the infinity of the radial coordinate, we get
\begin{eqnarray}
\rho(r)&\approx&\frac{16 \pi  c_1 m^2}{\sqrt{3} \kappa ^2 q^5}-\frac{16 c_1 m}{\kappa ^2 r^4}+\frac{192 c_1 m^2}{5 \kappa ^2 r^5}+\frac{6 m q^3}{\kappa ^2	r^6}+O\left(\frac{1}{r^7}\right),\\
p_r(r)&\approx&-\frac{16 \pi  c_1 m^2}{\sqrt{3} \kappa ^2 q^5}+\frac{16 c_1 m}{\kappa ^2 r^4}-\frac{192 c_1 m^2}{5 \kappa ^2 r^5}-\frac{6 m q^3}{\kappa ^2	r^6}+O\left(\frac{1}{r^7}\right),\\
p_t(r)&\approx&-\frac{16 \pi  c_1 m^2}{\sqrt{3} \kappa ^2 q^5}-\frac{16 c_1 m}{\kappa ^2 r^4}+\frac{288 c_1 m^2}{5 \kappa ^2 r^5}+\frac{12 m	q^3}{\kappa ^2 r^6}+O\left(\frac{1}{r^7}\right).
\end{eqnarray}
The term that is constant in the infinity comes from $f(G)$ gravity and it is also responsible for the isotropy in this region. The effective quantities are the same from general relativity, Eqs. \eqref{eq3}, where we have isotropy in the black hole center and the effective energy density and the pressures are zero in the infinity. We may consider that the effective energy density is related to the radial, tangential and total pressures by equations of states as $p_r^{(eff)}=\omega_r \rho^{(eff)}$, $p_t^{(eff)}=\omega_t \rho^{(eff)}$ and $\left(p_r^{(eff)}+2p_t^{(eff)}\right)=\omega_{tot}\rho^{(eff)} $. From $\omega_r$, $\omega_t$ and $\omega_{tot}$ we may extract informations about the energy conditions. If the effective energy density is positive, the case $\omega_{r,t}<-1$ implies in the violation of $WEC_{1,2}$, however, to $\omega_{r,t}>1$, $DEC_{2,3}$ is violated and to $\omega_{tot}<-1$, $SEC$ is violated. In Fig. \ref{ws} we plot the behavior of these parameters. We may see that $\omega_r=-1$, $-1<\omega_t<2$ and $-3<\omega_{tot}<3$, so that, $DEC_3$ is violated outside the event horizon, while $SEC$ is violated inside.

As we have an electric source, it is interesting to use the $H(P)$ formalism, that is a dual description of the same physical problem (for more details see \ref{ap1}). In this formalism, we don't work with the electric field but with the induction field \eqref{P} \cite{Burinskii}. In linear electrodynamics, we have $P^{\mu\nu}=F^{\mu\nu}$ but here they are different by the multiplicative term $\partial_IL$. In the nonlinear case $F^{10}$ does not behave as Maxwell, however, if we consider the displacement vector, inserting \eqref{ef} in \eqref{P}, we find
\begin{equation}
P^{10}=\frac{q}{r^2},
\end{equation}
which is the same result from the linear electrodynamics. Since we have $L$ as a function of the radial coordinate, we can also find an analytical expression for $L(P)$. From \eqref{PP}, we find an expression for $r(P)$, and with that, $L(P)$ is given by
\begin{eqnarray}
L(P)&=&-\frac{12 \sqrt[4]{2} \alpha  \beta  \sqrt{-P} P
	\left(2^{3/4} \beta ^2 \sqrt{-P} P+\beta  (-P)^{3/4} q^{3/2}+\sqrt[4]{2} q^3\right)}{\kappa ^2 \left( \beta  (-2P)^{3/4}+ q^{3/2}\right)^4}\nonumber\\
&+&\frac{\alpha  c_1 }{3 \beta ^{5/3} \kappa ^2}\left(4 \alpha  \left(\ln \left(\beta ^{2/3}-\frac{\sqrt[3]{\beta } \sqrt{q}}{\sqrt[4]{-2P} }+\frac{q}{\sqrt{-2P}}\right)-2 \ln	\left(\sqrt[3]{\beta }+\frac{\sqrt{q}}{ \sqrt[4]{-2P}}\right)+2 \sqrt{3} \tan ^{-1}\left(\frac{1}{\sqrt{3}}-\frac{2^{3/4} \sqrt{q}}{\sqrt[3]{\beta}\sqrt{3}(-P)^{1/4}}\right)\right)\right.\nonumber\\
&-&\frac{12 \sqrt[4]{2} \beta ^{2/3} \sqrt{-P} }{\sqrt{q} \left(\beta  (-2P)^{3/4}+ q^{3/2}\right)^4}\left(-48 \sqrt[4]{2} \beta ^3 P^2 q^{3/2}+16 \beta ^4 (-P)^{11/4}+48\ 2^{3/4} \alpha 
\beta ^2 (-P)^{3/2} q^{5/2}+2\ 2^{3/4} \beta  \sqrt{-P} q^{9/2}\right.\nonumber\\
&+&\left.\left.24 \sqrt{2} \beta ^2 \sqrt[4]{-P} P q^3-8 \sqrt{2} \alpha  \beta ^3 (-P)^{9/4} q+\sqrt[4]{2} \alpha 
		q^{11/2}\right)\right).
\end{eqnarray}
With that, the nonlinearity of the electromagnetic theory is clear. We comment on the model \eqref{sol2} and \eqref{sol3} with $f(G)$ theory in \ref{subsec8}. In that appendix, we obtain the analytical expressions for the $f_G(G)$ function and the electromagnetic quantities, considering \eqref{sol2}. For the solution \eqref{sol3}, we find the electromagnetic quantities, however, it is not possible to write $f_G(G)$ in a closed form. 

\section{Conclusion}\label{sec5}
In this work, we analyzed the existence of regular multihorizon black holes in general relativity and in some alternative theories of gravity. The method presented with general relativity was developed earlier in \cite{Odintsov}, however, we fixed some misprints which influenced the energy conditions. The energy density is positive in the model \eqref{sol1}, however, due to the presence of the cosmological constant, we can have negative energy density in the models \eqref{sol2} and \eqref{sol3}. We realized that the electromagnetic function $X(r)$ goes to zero in $r\rightarrow \infty$ for the solution that is asymptotically flat but tends to a constant when we consider the solution that is asymptotically anti-de Sitter and the solution with three horizons. The strong energy condition is violated inside the black hole for all solutions and outside for those that are not asymptotically Minkowski.

In modified theories, in order to satisfy all the field equations, we did not fixed the $f(R)$ and $f(G)$. Since we imposed the solutions that we wanted to find, we got the correspondent $f(R)$ and $f(G)$ theories and the modifications in the electromagnetic sector. Due to the coupled with the $f(R)$ gravity, $X(r)$ diverges in infinity of the radial coordinate for all solutions, while in $f(G)$ gravity, it diverges only for those that are not asymptotically flat solutions. It means that, in $f(G)$ theory, the electric field is well behaved in all spacetime for the asymptotically flat solution, while it diverges in $f(R)$ theory. Since $f_R$ and $f_G$ are linear in $r$, we have a divergence in infinity, however, $f(R)$ and $f(G)$, which are the functions that appear in the action, do not present divergence. Because of the symmetry $a=-b$, the energy conditions will be the same as in general relativity.

Since $G$ is topologically invariant in four dimensions, the authors of \cite{Odintsov} needed to consider five dimensions to find modifications in the field equations. Here, as we consider a nonlinear function of $G$ in the action, we to consider high dimensions to find new regular solutions.

As we considered electric sources, we showed the nonlinear behavior of the electromagnetic theory numerically by a parametric plot of $L(I)$. In \cite{Odintsov}, the authors used auxiliary fields, $B$ and $C$, to find an analytical expression for what they called $L_{BCX}$, however, the nonlinearity of this Lagrangian is not so clear, so that, we consider the scalar $P$, that is related to the electromagnetic scalar, to represent analytically the electromagnetic Lagrangian in \ref{ap1}.

We can also analyze the consequences to $a\neq -b$ in alternative theories of gravity. As this symmetry modifies the line element, differences in the energy conditions and in the curvature invariants are expected to appear, so some choices can give solutions that are not regular.
\vspace{1cm}

{\bf Acknowledgments}:The authors would like to thank Shin'ichi Nojiri and Sergei D. Odintsov. M. E. R.  thanks Conselho Nacional de Desenvolvimento Cient\'ifico e Tecnol\'ogico - CNPq, Brazil  for partial financial support. This study was financed in part by the Coordenação de Aperfeiçoamento de Pessoal de Nível Superior - Brasil (CAPES) - Finance Code 001.


\appendix
\section{$H(P)$ formalism}\label{ap1}
As we already mentioned, for the electric sources it is not possible to construct an analytical closed form for $r(I)$. However, it is still possible to find an expression for the electric Lagrangian in terms of a new scalar $P=P^{\mu\nu}P_{\mu\nu}$, where the field $P_{\mu\nu}$ is defined as
\begin{equation}
P_{\mu\nu}=\partial_IL F_{\mu\nu}.
\label{P}
\end{equation}
As the nonzero components of $F_{\mu\nu}$ are proportional to $\left(\partial_IL\right)^{-1}$, the scalar $P$ is given by
\begin{equation}
P=-\frac{q^2}{2r^4}.
\label{PP}
\end{equation}
Using this relation, we can find an expression for $r(P)$. Moreover, we can make the Legendre transformation of $L$, which results in
\begin{equation}
\mathcal{H}=2 I\partial_I L-L.
\end{equation}
For a spherically symmetric line element such as \eqref{ele}, $\mathcal{H}$, in terms of the radial coordinate, is
\begin{equation}
\mathcal{H}(r)=\frac{e^{a} \left(r a'+1\right)-1}{\kappa ^2 r^2}.
\end{equation}
Since we have $\mathcal{H}(r)$ and $r(P)$, from \eqref{P}, we may find the function $L(P)$ using an inverse Legendre transformation as
\begin{equation}
L(P)=2 P \partial_P \mathcal{H}-\mathcal{H}.
\end{equation}
Considering the models \eqref{sol1}, \eqref{sol2} and \eqref{sol3}, $L(P)$ is
\begin{eqnarray}
L_1(P)&=&\frac{24 \alpha  \beta  (-P)^{3/2} \left(\beta  (-P)^{3/4}-\sqrt[4]{2} q^{3/2}\right)}{\kappa ^2 \left(2 \beta  (-P)^{3/4}+\sqrt[4]{2} q^{3/2}\right)^3},\\
L_2(P)&=&\frac{24 P^2 r_1^2 r_2^2 \left(r_1^2+r_1 r_2+r_2^2\right)-6\sqrt{-2P} q^3+36 P q^2 r_1 r_2+4 \sqrt{-2P} P q
	r_1 r_2 \left(r_1^2+11 r_1 r_2+r_2^2\right)}{\kappa ^2 \sqrt{-P} r_1 r_2 \left(2 \sqrt{-P} r_1 r_2+\sqrt{2}	q\right)^3},\\
L_3(P)&=&-\frac{8 \sqrt{-P}}{\left(\sqrt{2} l q+2 \sqrt{-P} \gamma\right)^{-5}}\left\{\left(-\sqrt{-2P} q^2 l^4+8 P q \gamma
	l^3+\left(\sqrt{-2P} \omega q^2-4 P \delta q+30 \sqrt{-2P} P \gamma^2\right) l^2\right.\right.\nonumber\\
&+&\left.\left.2 \gamma \left(\sqrt{-2P} q^2-8 P\omega q+10 \sqrt{-2P} P \delta\right) l+2 P
	\gamma^2 \left(5 \sqrt{-2P} \omega-14 q\right)\right)^2 \kappa ^2\right\}^{-1} \left(q^6 \left(2 \sqrt{-2P} P \delta q-3 \sqrt{-2P} q^3\right.\right.\nonumber\\
&+&\left.24 P^2	\gamma^2\right) l^{11}+2 P q^5 \gamma \left(39 q^3+2 \sqrt{-2P} \omega	q^2-2 P \delta q+96 \sqrt{-2P} P \gamma^2\right) l^{10}-2 q^4
	\left(12 P \delta	q^4-3 \sqrt{-2P} \omega q^5\right.\nonumber\\
&+&\left.\sqrt{-2P} P \left(2 \omega\delta -291 \gamma^2\right) q^3-8 P^2 \left(\delta^2-5\omega\gamma^2\right) q^2+18 \sqrt{2} (-P)^{5/2} \gamma^2 \delta q+1080 P^3 \gamma^4\right) l^9\nonumber\\
&-&4 q^3 \gamma \left(-3	\sqrt{-2P} q^6+51 P \omega q^5+2 \sqrt{-2P} P \left(\omega^2+13\delta\right) q^4-9 P^2 \left(2\delta\omega -151\gamma^2\right) q^3\right.\nonumber\\
&+&\left.2 \sqrt{2} (-P)^{5/2} \left(16\delta^2-81\gamma^2\omega\right) q^2+526 P^3 \gamma^2 \delta q-1344 \sqrt{2} (-P)^{7/2} \gamma^4\right) l^8-q^2 \left(3 \sqrt{-2P} \omega^2 q^7-17184 P^4 \gamma^6\right.\nonumber\\
&-&24 P \left(\delta\omega-16\gamma^2\right) q^6+2 \sqrt{2}	(-P)^{3/2} \left(\delta\omega^2+12\delta^2-686\omega\gamma^2\right) q^5+8 P^2 \left(2\omega\delta^2-19\gamma^2\omega^2-14\gamma^2\delta\right) q^4\nonumber\\
&+&\left.4 \sqrt{2} (-P)^{5/2} \left(4\delta^3+4445\gamma^4+110\gamma^2\delta\omega\right) q^3+32 P^3 \gamma^2 \left(3\delta^2-71 \gamma^2 \omega  \right) q^2+7720 \sqrt{2} (-P)^{5/2} P \gamma^4 \delta	 q\right) l^7\nonumber\\
&+&2 q \gamma \left(-6\sqrt{-2P} \omega q^8+3 P \left(21 \omega^2+8\delta\right) q^7+2 \sqrt{-2P} P \left(\omega^3+50 \left(\omega\delta-3\gamma^2\right) -504 \gamma^2\right) q^6\right.\nonumber\\
&+&2 P^2 \left(52\delta^2- 17\omega^2\delta+2774 \gamma^2 \omega\right) q^5+8 \sqrt{2} (-P)^{5/2} \left(2\delta^2\omega-336 \gamma^2\delta+ 13\gamma^2\omega^2 \right) q^4+30504 P^4 \gamma^4 \delta q\nonumber\\
&+&\left.4 P^3 \left( 366  \delta\omega\gamma^2 +9321 \gamma^4-68\delta^3\right) q^3+8 \sqrt{2}(-P)^{5/2} P \gamma^2 \left(43 \gamma^2 \omega +184\delta^2 \right) q^2-5760 \sqrt{2} (-P)^{9/2} \gamma^6\right) l^6\nonumber\\
&+&2 \gamma^2 \left(-6  \sqrt{-2P} q^9+240 P \omega q^8+\sqrt{-2P} P \left(443\omega^2+172 \delta\right) q^7-24 P^2 \left(2\omega^3-15 \delta\omega - 444 \gamma^2\right)q^6\right.\nonumber\\
&+&2 \sqrt{2} (-P)^{5/2} \left(71\delta\omega^2+ 6138 \gamma^2\omega+208\delta^2  \right) q^5-8 P^3 \left(198\delta^2\omega-97\gamma^2\omega^2- 3070\gamma^2\delta \right) q^4\nonumber\\
&+&12 \sqrt{2} (-P)^{5/2} P \left(120\delta^3+1134 \delta\omega \gamma^2 +4067 \gamma^4 \right)q^3+32 P^4 \gamma^2 \left(691\delta^2+1101 \gamma^2 \omega \right) q^2+5160 \sqrt{2} (-P)^{9/2} \gamma^4 \omega q\nonumber\\
&+&\left.21600 P^5 \gamma^6\right) l^5+4 P\gamma^3\left(114 q^8+826 \sqrt{-2P} \omega q^7-P \left(1207\omega^2-308 \delta \right) q^6\right.\nonumber\\
&+&2 \sqrt{-2P} P \left(15\omega^3 +1094 \delta\omega+6618 \gamma^2 \right) q^5-2 P^2 \left(1031\delta\omega^2 + 168\delta^2+6338 \gamma^2\omega \right) q^4\nonumber\\
&+&8 \sqrt{2} (-P)^{5/2} \left(3929 \gamma^2\delta+829 \gamma^2\omega^2+ 448\delta^2\omega\right)	q^3+4 P^3 \left( 488\delta^3 -3693 \gamma^4+3430 \delta\omega\gamma^2\right) q^2\nonumber\\
&+&\left.120 \sqrt{2} (-P)^{5/2} P \gamma^2 \left(56\delta^2+115\gamma^2\omega\right) q+19800 P^4 \gamma^4\delta\right) l^4+4 P \gamma^4 \left(780 \sqrt{-2P} q^7-4784P \omega q^6\right.\nonumber\\
&+&\sqrt{-2P} P \left(3720\delta-881\omega^2\right) q^5-8 P^2 \left(188\omega^3 +1703\delta\omega +4656\gamma^2\right) q^4\nonumber\\
&+&2\sqrt{2} (-P)^{5/2} \left(1883\delta\omega^2+ 4364\delta^2+7030 \gamma^2\omega\right) q^3+8 P^3 \left(82\delta^2\omega-487\gamma^2\omega^2-3990\gamma^2\delta\right) q^2\nonumber\\
&+&\left.40 \sqrt{2} (-P)^{5/2} P\left(62\delta^3+152\gamma^4 +543 \delta\omega\gamma^2\right) q+2400 P^4 \gamma^2 \left(5\delta^2+3 \gamma^2 \omega\right)\right) l^3+8 P^2 \gamma^5 \left(-2436 q^6\right.\nonumber\\
&-&2898 \sqrt{-2P} \omega q^5-P \left(3643 \omega^2+10544\delta\right) q^4+6 \sqrt{-2P} P \left(25 \omega^2 +290 \delta\omega -1414 \gamma^2 \right) q^3\nonumber\\
&-&2 P^2 \left(3548\delta^2 +1387\delta\omega^2+8574 \gamma^2\omega\right) q^2+120 \sqrt{2} (-P)^{5/2} \left(34\delta^2\omega+37\gamma^2\omega^2 +49\gamma^2\delta\right) q\nonumber\\
&+&\left.600 P^3 \delta \left(2\delta^2 +7 \gamma^2 \omega\right)\right) l^2+8 P^2 \gamma^6 \left(-3822 \sqrt{-2P} q^5-2464 P \omega q^4+\sqrt{2} (-P)^{3/2} \left(1775\left(\omega^2-2\delta\right) +10242\delta\right) q^3\right.\nonumber\\
&-&\left.8 P^2 \left(150 \omega^3+1553\delta\omega+294 \gamma^2 \right) q^2+210 \sqrt{2} (-P)^{5/2} \left(13 \delta\omega^2+8\delta^2+8 \gamma^2\omega\right) q+600 P^3 \omega\left(2\delta^2+\omega\gamma^2\right)\right) l\nonumber\\
&+&16 P^3 \gamma^7 \left(2058 q^4-2450 \sqrt{-2P} \omega q^3-7 P\left(275 \omega^2+84 \delta\right) q^2+10 \sqrt{-2P} P\omega \left(25\omega^2+42 \delta\right) q\right.\nonumber\\
&+&\left.\left.150 P^2 \omega^2 \delta\right)\right).
\end{eqnarray}
With that, the electromagnetic Lagrangians associated with \eqref{sol1}, \eqref{sol2} and \eqref{sol3} are not linear in the scalar $P$ and do not behave as Maxwell to $P\approx 0$.
\section{Corrections from $f(R)$ theory}\label{ap2}
In this appendix, we will write the expressions for $\rho(r)$, $L(r)$ and $X(r)$ with the corrections from $f(R)$ gravity for the models \eqref{sol2} and \eqref{sol3}.
\subsection{Example in the anti-de Sitter background}\label{subsec6}
Inverting \eqref{cs2}, we find the expression for $r(G)$ and with
\begin{eqnarray}
\alpha_2(R)&=&3 \sqrt{3} \left(r_1^2	r_2^2 (r_1+r_2)^4 (R r_1 r_2-12)^2 \left(\left(2 R \left(54 R r_2^2-17\right) r_2^2+15\right) r_1^4+4
r_2 \left(117-665 R r_2^2\right) r_1^3\right.\right.\nonumber\\
&+&\left.\left.2 r_2^2 \left(8229-17 R r_2^2\right) r_1^2+468 r_2^3 r_1+15r_2^4\right)\right)^{1/2},\\
\alpha_3(R)&=&\left(9 R \left(6 R r_{2}^2-1\right) r_{2}^2+4\right) r_{1}^4,\\
\alpha_4(R)&=&(r_1+r_2)^2 \left(\alpha_3(R)+2 r_{2} \left(62-657 Rr_{2}^2\right) r_{1}^3+\left(8016-9 R r_{2}^2\right) r^2_2r_1^2+124 r_{2}^3 r_{1}+4 r_{2}^4\right),
\end{eqnarray}
$f_R(R)$ is
\begin{eqnarray}
f_R(R)&=&c_0+\frac{c_1\left(\alpha_4(R)-\alpha_2(R)\right)^{-1}}{\sqrt{6}(R r_1 r_2-12)} \left[-2 \left(3 R r_2^2+2\right) \alpha_3(R) r_1^4-8 r_2 \left(9 R\left(9 R r_2^2 \left(R r_2^2-17\right)-55\right) r_2^2+34\right) r_1^7\right.\nonumber\\
&+&\left(4 i \left(i+\sqrt{3}\right) \sqrt[3]{2\alpha_2(R)-2 \alpha_4(R)}-r_2^2 \left(18 R \left(3 R \left(6 R r_2^4-414 r_2^2+\left(1-i \sqrt{3}\right) \sqrt[3]{2\alpha_2(R)-2\alpha_4(R)}\right)\right.\right.\right.\nonumber\\
&+&\left.\left.\left.6902\right) r_2^2+9 i \left(i+\sqrt{3}\right) R \sqrt[3]{2\alpha_2(R)-2\alpha_4(R)}+24640\right)\right) r_1^6+4 r_2 \left(\left(3 R
\left(918 R r_2^2\right.\right.\right.\nonumber\\
&+&\left.\left.9 i \left(i+\sqrt{3}\right) R \sqrt[3]{2\alpha_2(R)-2 \alpha_4(R)}-21364\right) r_2^2+333 \left(1-i \sqrt{3}\right) R \sqrt[3]{2\alpha_2(R)-2 \alpha_4(R)}+115876\right) r_2^2\nonumber\\
&+&\left.33 i \left(i+\sqrt{3}\right) \sqrt[3]{2\alpha_2(R)-2 \alpha_4(R)}\right) r_1^5+\left(-162 R^2 r_2^8+18 i R \left(3
			\left(i+\sqrt{3}\right) \sqrt[3]{\alpha_2(R)-2 \alpha_4(R)} R+6902 i\right) r_2^6\right.\nonumber\\
&+&2 \left(1323 \left(1-i \sqrt{3}\right) \sqrt[3]{2\alpha_2(R)-2 \alpha_4(R)} R+487888\right)
			r_2^4-3 i \sqrt[3]{2\alpha_2(R)-2 \alpha_4(R)}\times\nonumber\\
&&\left. \left(\left(-i+\sqrt{3}\right) R \sqrt[3]{2\alpha_2(R)-2 \alpha_4(R)}-2756 \left(i+\sqrt{3}\right)\right) r_2^2+2 \left(1+i \sqrt{3}\right) \left(2\alpha_2(R)-2\alpha_4(R)\right)^{2/3}\right) r_1^4\nonumber\\
&+&2 r_2 \left(1980 R r_2^6+2 \left(333 \left(1-i \sqrt{3}\right) \sqrt[3]{2\alpha_2(R)-2 \alpha_4(R)} R+115876\right)
			r_2^4-i \sqrt[3]{2\alpha_2(R)-2 \alpha_4(R)}\times\right.\nonumber\\
&&\left. \left(3 \left(-i+\sqrt{3}\right) R \sqrt[3]{\alpha_2(R)-2 \alpha_4(R)}-8140 \left(i+\sqrt{3}\right)\right) r_2^2+22 \left(1+i
			\sqrt{3}\right) \left(\alpha_2(R)-2 \alpha_4(R)\right)^{2/3}\right) r_1^3\nonumber\\
&+&\left(12 R r_2^8+\left(9 \left(1-i \sqrt{3}\right) R \sqrt[3]{2\alpha_2(R)-2 \alpha_4(R)}-24640\right) r_2^6-3 i \sqrt[3]{\alpha_2(R)-2 \alpha_4(R)}\times\right.\nonumber\\
&& \left(\left(-i+\sqrt{3}\right) R \sqrt[3]{\alpha_2(R)-2 \alpha_4(R)}-2756
			\left(i+\sqrt{3}\right)\right) r_2^4+6 \left(\alpha_2(R) R\right.\nonumber\\
&+&\left.\left.14 \left(1+i \sqrt{3}\right) \left(2\alpha_2(R)-2 \alpha_4(R)\right)^{2/3}\right) r_2^2+4\alpha_2(R)\right) r_1^2-4 \left(68 r_2^7+33 \left(1-i
			\sqrt{3}\right) \sqrt[3]{2\alpha_2(R)-2 \alpha_4(R)} r_2^5\right.\nonumber\\
&-&\left.11 i \left(-i+\sqrt{3}\right) \left(2\alpha_2(R)-2 \alpha_4(R)\right)^{2/3} r_2^3+16\alpha_2(R) r_2\right)
			r_1-16 r_2^8+2 \left(1+i \sqrt{3}\right) r_2^4 \left(2\alpha_2(R)-2 \alpha_4(R)\right)^{2/3}\nonumber\\
&+&\left.4 r_2^2 \alpha_2(R)+4 i \left(i+\sqrt{3}\right)
			r_2^6 \sqrt[3]{2\alpha_2(R)-2 \alpha_4(R)}+ \left(-\sqrt{3} i+1\right) \alpha_2(R) \sqrt[3]{2\alpha_2(R)-2 \alpha_4(R)}\right]^{1/2}.
\end{eqnarray}
Integrating this expression with respect to the curvature scalar we obtain the $f(R)$ function. The corrections to $\rho$, $L$ and $X$ are
\begin{eqnarray}
\rho(r)&=&\frac{c_0 \left(-3 r^4+r^2 \left(r_1^2-4 r_1 r_2+r_2^2\right)+3 r_1 r_2 \left(r_1^2+r_1 r_2+r_2^2\right)\right)}{\kappa
	^2 r_1 r_2 \left(r^2+r_1 r_2\right)^2}+\frac{c_1}{\kappa ^2 (r_1 r_2)^{3/2} } \left(-3 (r_1+r_2)^2\tan ^{-1}\left(\frac{r}{\sqrt{r_1r_2} }\right)\right.\nonumber\\
&+&\left.\frac{\sqrt{r_1r_2} \left(r^4 \left(r_1^2+r_2^2\right)+r^2
	r_1 r_2 \left(3 r_1^2+2 r_1 r_2+3 r_2^2\right)-2 r_1^3 r_2^3\right)}{r \left(r^2+r_1 r_2\right)^2}\right),\\
L(r)&=&\frac{c_0 \left(3 r^6+9 r^4 r_1 r_2+r^2 r_1 r_2 \left(r_1^2+11 r_1 r_2+r_2^2\right)-3 r_1^2 r_2^2
	\left(r_1^2+r_1 r_2+r_2^2\right)\right)}{\kappa ^2 r_1 r_2 \left(r^2+r_1 r_2\right)^3}\nonumber\\
&+&\frac{c_1 }{\kappa ^2 (r_1r_2)^{3/2}}\left(\frac{\sqrt{r_1r_2} \left(r^6 \left(r_1^2+3 r_1 r_2+r_2^2\right)+r^4 r_1 r_2 \left(2 r_1^2+7 r_1 r_2+2 r_2^2\right)-3 r^2
		r_1^2 r_2^2 \left(r_1^2+r_1 r_2+r_2^2\right)+r_1^4 r_2^4\right)}{r \left(r^2+r_1 r_2\right)^3}\right.\nonumber\\
&+&\left.3 (r_1+r_2)^2	\tan ^{-1}\left(\frac{r}{\sqrt{r_1} \sqrt{r_2}}\right)\right),\\
X(r)&=&\frac{c_0 r^4 (r_1+r_2)^2 \left(r^2+5 r_1 r_2\right)}{\kappa ^2 r_1 r_2 \left(r^2+r_1 r_2\right)^3}+\frac{c_1 r}{\kappa ^2 r_1 r_2 \left(r^2+r_1 r_2\right)^3} \left(r^6
	\left(2 r_1^2+3 r_1 r_2+2 r_2^2\right)+3 r^4 r_1 r_2 \left(2 r_1^2+3 r_1 r_2+2 r_2^2\right)\right.\nonumber\\
&-&\left.3 r^2 r_1^3r_2^3-r_1^4 r_2^4\right).
\end{eqnarray}
The electromagnetic energy density tends to a constant in infinity while $X(r)$ diverges due to the presence of $c_1$.

\subsection{Regular black hole with three horizons}
As naturally the expressions for the solution with three horizons are more complicated, we expect that the $f(R)$ function that generates this model is not simple. Choosing
\begin{eqnarray}
\alpha_5(R)&=&4 l^3+3 R \gamma l^2-4 \omega l+28 \gamma,\ \alpha_6(R)=-3 \gamma l^2+6\delta l-2l\gamma^2R+15 \gamma \omega,\nonumber\\
\alpha_7(R)&=&3 l^3-4 R \gamma l^2-3 \omega l-54 \gamma,\ \alpha_8(R)=l^3+2 R \gamma l^2-\omega l+22 \gamma,\\
\alpha_9(R)&=&12\delta+\gamma^2R+24 l \gamma,\nonumber
\end{eqnarray}
the $f_R$ function is
\begin{eqnarray}
f_R&=&c_0+\frac{c_1}{\sqrt{2}}\left\{-\frac{\alpha_8}{l \left(Rl^2+12\right)}-\left[\frac{\alpha_7 \left(l^3-\omega l-2 \gamma\right)}{3 l^2 \left(R l^2+12\right)^2}+\frac{\sqrt[3]{2}}{3 l^2 \left(R l^2+12\right)} \left(4\gamma^3 \alpha_5^3-36\gamma^3 l^2 \alpha_5\left(R l^2+12\right)  \alpha_9\right.\right.\right.\nonumber\\
&+&\left.27 \gamma^2l^2 \left(R l^2+12\right)\alpha_6^2+18	l \gamma^2 \alpha_5 \alpha_8 \alpha_6+27 l^2 \gamma^2 \alpha_8^2\alpha_9+\left(\gamma^3 \left(\gamma \left(4\gamma \alpha_5^3-36 l^2\gamma \left(R l^2+12\right)\alpha_5\alpha_9+27 l^2 \alpha_8^2 \alpha_9\right. \right.\right.\right.\nonumber\\
&+&\left.18l \alpha_8 \alpha_6 \alpha_5+27 l^2 \left(R l^2+12\right) \alpha_6^2\right)^2 -\left.\left.\left.16 \left(3\gamma l^2 \left(R l^2+12\right)  \alpha_9+3 \alpha_8 \alpha_6 l+\gamma \alpha_5^2\right)^3\right)\right)^{1/2}\right)^{1/3}\nonumber\\
&+&\frac{2\ 2^{2/3} \gamma}{3 l^2 \left(R l^2+12\right)} \left(3 \left(R l^2+12\right) \gamma \alpha_9 l^2+3l\alpha_8 \alpha_6+\gamma \alpha_5^2\right)\left(4 \gamma^3 \alpha_5^3-36 l^2\gamma^3 \left(Rl^2+12\right) \alpha_5 \alpha_9+27 l^2\gamma^2 \left(R l^2+12\right) \alpha_6^2\right.\nonumber\\
&+&18 l \gamma^2\alpha_5 \alpha_8 \alpha_6+27 l^2 \gamma^2 \alpha_8^2 \alpha_9+\left(\gamma^3 \left(\gamma \left(4 \gamma \alpha_5^3+18 l \alpha_8 \alpha_6 \alpha_5+27 l^2 \alpha_8^2 \alpha_9+27 l^2 \left(Rl^2+12\right) \alpha_6^2\right.\right.\right.\nonumber\\
&-&\left.\left.\left.\left.\left.36 l^2\gamma \left(R l^2+12\right) \alpha_9\alpha_5\right)^2-16 \left(3\gamma \left(R l^2+12\right) \alpha_9 l^2+3 \alpha_8\alpha_6 l+\gamma \alpha_5^2\right)^3\right)\right)^{1/2}\right)^{-1/3}\right]^{1/2}+\left[\frac{2 \alpha_7 \left(l^3-\omega l-2 \gamma\right)}{3 l^2 \left(R l^2+12\right)^2}\right.\nonumber\\
&-&\left.\frac{\sqrt[3]{2}}{3 l^2 \left(R l^2+12\right)} \left(4\gamma^3\alpha_5^3-36 l^2\gamma^3 \left(R l^2+12\right) \right.\right.\alpha_5\alpha_9+27 l^2 \left(R l^2+12\right) \gamma^2 \alpha_6^2+18 l \gamma^2 \alpha_5 \alpha_8 \alpha_6+27	l^2 \gamma^2 \alpha_8^2\alpha_9\nonumber\\
&+&\left(\gamma^3 \left(\gamma \left(18 l \alpha_8 \alpha_6 \alpha_5\right.\right.-36 l^2 \left(R l^2+12\right) \gamma \alpha_9\alpha_5+4 \gamma \alpha_5^3+27 l^2 \left(Rl^2+12\right) \alpha_6^2+27 l^2 \alpha_8^2 \alpha_9\right)^2\nonumber\\
&-&\left.\left.\left.16 \left(3 \left(R l^2+12\right) \gamma\alpha_9 l^2+3 \alpha_8 \alpha_6 l+\gamma \alpha_5^2\right)^3\right)\right)^{1/2}\right)^{1/3}-\frac{2\ 2^{2/3} \gamma}{3 l^2 \left(R l^2+12\right) } \left(3 \left(R l^2+12\right) \gamma\alpha_9 l^2+3 \alpha_8 \alpha_6 l+\gamma \alpha_5^2\right)\times\nonumber\\
&&\left(4 \gamma^3 \alpha_5^3+18 l \gamma^2 \alpha_5\alpha_8 \alpha_6\right.+27 l^2 \left(R l^2+12\right) \gamma^2 \alpha_6^2+27 l^2 \gamma^2 \alpha_8^2\alpha_9+\left(\gamma^3 \left(\gamma \left(18	l \alpha_8 \alpha_6 \alpha_5-36 l^2 \left(R l^2+12\right) \gamma \alpha_9 \alpha_5\right.\right.\right.\nonumber\\
&+&\left.\left.\left.\left.27 l^2 \left(R l^2+12\right) \alpha_6^2+4\gamma \alpha_5^3+27 l^2 \alpha_8^2 \alpha_9\right)^2-16 \left(3l^2\gamma \left(R l^2+12\right) \alpha_9 +3 \alpha_8 \alpha_6 l+\gamma \alpha_5^2\right)^3\right)\right)^{1/2}\right)^{-1/3}\nonumber\\
&-&\frac{2}{l^3 \left(R l^2+12\right)^3} \left(-\alpha_8^2+2 \left(R l^2+12\right) \gamma \alpha_5 \alpha_8+2 l \left(R l^2+12\right)^2 \gamma \alpha_6\right)\left(\frac{\alpha_7\left(l^3-\omega l-2 \gamma\right)}{3 l^2 \left(R l^2+12\right)^2}+\frac{\sqrt[3]{2}}{3 l^2 \left(R l^2+12\right)} \left(4 \gamma^3  \alpha_5^3\right.\right.\nonumber\\
&-&36 l^2 \left(R l^2+12\right) \gamma^3 \alpha_5 \alpha_9+27 l^2 \left(R l^2+12\right) \gamma^2 \alpha_6^2+18 l \gamma^2 \alpha_5\alpha_6\alpha_8 +27l^2 \gamma^2 \alpha_8^2\alpha_9+\left(\gamma^3 \left(\gamma \left(4 \gamma \alpha_5^3+18 l \alpha_8 \alpha_6 \alpha_5\right.\right.\right.\nonumber\\
&-&\left.\left.\left.\left.36 l^2 \left(R l^2+12\right) \gamma \alpha_9\alpha_5+27 l^2 \left(Rl^2+12\right) \alpha_6^2+27 l^2 \alpha_8^2 \alpha_9\right)^2-16 \left(3 \left(R l^2+12\right) \gamma\alpha_9 l^2+3 \alpha_8 \alpha_6 l+\gamma \alpha_5^2\right)^3\right)\right)^{1/2}\right)^{1/3}\nonumber\\
&+&\frac{2\ 2^{2/3} \gamma}{3 l^2 \left(R l^2+12\right) } \left(3 \left(R l^2+12\right) \gamma \alpha_9 l^2+3 \alpha_8\alpha_6 l+\gamma \alpha_5^2\right)\left(4 \gamma^3 \alpha_5^3 -36 l^2	\left(R l^2+12\right) \gamma^3 \alpha_5 \alpha_9+27 l^2 \left(R l^2+12\right) \gamma^2 \alpha_6^2\right.\nonumber\\
&+&18l \gamma^2 \alpha_5	\alpha_8 \alpha_6+27 l^2 \gamma^2 \alpha_8^2 \alpha_9+\left(\gamma^3 \left(\gamma \left(4\gamma \alpha_5^3+18l \alpha_8 \alpha_6 \alpha_5-36 l^2 \left(R l^2+12\right) \gamma \alpha_9 \alpha_5+27 l^2 \left(R l^2+12\right) \alpha_6^2\right.\right.\right.\nonumber\\
&+&\left.\left.\left.\left.\left.\left.\left.27 l^2 \alpha_8^2\alpha_9\right)^2-16 \left(3 \left(R l^2+12\right) \gamma \alpha_9 l^2+3 \alpha_8 \alpha_6 l+\gamma \alpha_5^2\right)^3\right)\right)^{1/2}\right)^{-1/3}\right)^{-1/2}\right]^{1/2}\right\}^{1/2}.
\end{eqnarray}

The corrections to the electromagnetic quantities are
\begin{eqnarray}
\rho(r)&=&\frac{c_0 }{\kappa ^2 \left(l r^2+\gamma\right)^3}\left(l^3 r^4+3 l^2 r^2 \gamma+l \left(3 r^6-r^4 \omega-r^2 \delta+6 \omega^2\right)+\gamma \left(7 r^4-5 r^2\omega+3 \delta\right)\right)\nonumber\\
&+&\frac{c_1 }{2 \kappa ^2l^{5/2} \gamma^{3/2}}\left(-3 \left(3 l^3 \gamma\right.+l^2 \delta-l \gamma \omega-3 \gamma^2\right) \tan ^{-1}\left(\sqrt{\frac{l}{\gamma}}r\right)-\frac{\sqrt{l\gamma} }{r \left(l r^2+\gamma\right)^3}\left(5 l^5 r^6 \gamma+3 l^4 r^4 \left(r^2 \delta+4 \gamma^2\right)\right.\nonumber\\
&+&\left.l^3 r^2 \gamma \left(5 r^4
		\omega+8 r^2 \delta+3 \gamma^2\right)+l^2 \gamma^2 \left(7 r^6+16 r^4 \omega-3 r^2 \delta+\left.4 \gamma^2\right)+3 l r^2 \gamma^3 \left(8r^2+\omega\right)+9 r^2 \gamma^4\right)\right),\\
L(r)&=&-\frac{c_0}{\kappa ^2 \left(l r^2+\gamma\right)^4} \left(l^2 \left(3 r^8+r^4 \delta-6 r^2 \gamma^2\right)+2
	l \gamma \left(6 r^6+r^4 \omega-4 r^2 \delta+3 \gamma^2\right)+\gamma^2 \left(21 r^4-10 r^2 \omega+3\delta\right)\right)\nonumber\\
&+&\frac{c_1}{2 \kappa ^2 \left(l r^2+\gamma\right)^4} \left(7 l^4 r^7+l^3 r^5\left(\frac{3r^2}{\gamma}\delta+25 \gamma\right)+l^2 r^3 \left(r^4 \omega
+11 r^2 \delta+39 \gamma^2\right)+\frac{9 r \gamma^4}{l^2}+\frac{3 r \gamma^3 \left(11	r^2+\omega\right)}{l}\right.\nonumber\\
&+&\frac{3\left(3 l^3 \gamma+l^2 \delta-l \gamma \omega-3 \gamma^2\right) \left(l r^2+\gamma\right)^4 \tan ^{-1}\left(\sqrt{\frac{l}{\gamma}}r\right)}{l^{5/2} \gamma^{3/2}}+l r \gamma \left(-r^6+5 r^4 \omega+29 r^2 \delta-\gamma^2\right)\nonumber\\
&+&\left.\frac{\gamma^2 \left(-r^6+31 r^4\omega-3 r^2 \delta+2 \gamma^2\right)}{r}\right),\\
X(r)&=&\frac{c_0 r^4 }{\kappa ^2 \left(l r^2+\gamma\right)^4}\left(l^4 r^4+4 l^3 r^2 \gamma+l^2 \left(-r^4\omega-2 r^2 \delta+15 \gamma^2\right)-2 l \gamma \left(r^4+4 r^2
	\omega-5 \delta\right)+\gamma^2 \left(5	\omega-14 r^2\right)\right)\nonumber\\
&+&\frac{c_1 r}{\kappa ^2 \left(l r^2+\gamma\right)^4} \left(l^4 r^8+4 l^3 r^6 \gamma-2 l^2 \left(r^8 \omega-6 r^4 \gamma^2\right)-4 l r^2 \gamma \left(r^6+2 r^4 \omega-3 r^2 \delta+\gamma^2\right)\right.\nonumber\\
&-&\left.\gamma^2 \left(16 r^6-6 r^4
	\omega+\gamma^2\right)\right).
\end{eqnarray}
We get the same conclusions as the anti-de Sitter model.
\section{Corrections from $f(G)$ theory}\label{ap3}
This appendix is dedicated to the corrections from $f(G)$ gravity in $\rho(r)$, $X(r)$ and $L(r)$.

\subsection{Example in the anti-de Sitter background}\label{subsec8}
From \eqref{G2}, we can find $r(G)$, and using
\begin{eqnarray}
\beta_1(G)&=&24-G r_1^2 r_2^2,\\
\beta_2(G)&=&243 G r_2^4 r_1^{12}+3024 G r_2^5 r_1^{11}+108 r_2^2 \left(127 G r_2^4-68\right) r_1^{10}+16 r_2^3 \left(1971 G r_2^4-5309\right) r_1^9\nonumber\\
&+&6 r_2^4 \left(6867 G r_2^4-62192\right) r_1^8+48 r_2^5 \left(657 G r_2^4-17617\right) r_1^7+4 r_2^6 \left(3429 G r_2^4-275032\right) r_1^6\nonumber\\
&+&48 r_2^7 \left(63 Gr_2^4-17617\right) r_1^5+3 r_2^8 \left(81 G r_2^4-124384\right) r_1^4-84944 r_2^9 r_1^3-7344 r_2^{10} r_1^2,\\
\beta_3(G)&=&27 G^2 r_2^5 \left(256 G r_2^4+81\right) r_1^{13}+216 G^2 r_2^6 \left(128G r_2^4+267\right) r_1^{12}+36 G r_2^3 \left(1152 G^2 r_2^8-3307 G r_2^4-3096\right) r_1^{11}\nonumber\\
&+&72 \left(384G^3 r_2^{12}-12857 G^2 r_2^8-40212 G r_2^4+96\right) r_1^{10}+18 r_2 \left(384 G^3 r_2^{12}-83079 G^2r_2^8-383392 G r_2^4+88576\right) r_1^9\nonumber\\
&-&8 r_2^2 \left(115713 G^2 r_2^8+624716 G r_2^4-4689984\right) r_1^8-4r_2^3 \left(29763 G^2 r_2^8+523184 G r_2^4-35735424\right) r_1^7\nonumber\\
&+&8 r_2^4 \left(7209 G^2 r_2^8-624716 Gr_2^4+34942560\right) r_1^6+3 r_2^5 \left(729 G^2 r_2^8-2300352 G r_2^4+116347904\right) r_1^5\nonumber\\
&-&96 r_2^6 \left(30159 Gr_2^4-2911880\right) r_1^4-96 r_2^7 \left(1161 G r_2^4-1488976\right) r_1^3+37519872 r_2^8 r_1^2+1594368 r_2^9 r_1\nonumber\\
&+&6912r_2^{10},\\
\beta_4(G)&=&\sqrt[3]{\beta_2(G)+3 \sqrt{3}\sqrt{r_1^3 r_2^3 (r_1+r_2)^8 \beta_3(G)}},\\
\beta_5(G)&=&\frac{8 \sqrt[3]{2} r_1 r_2 \beta_1(G) \left(9 r_2^2+29 r_1 r_2+9r_1^2-9 \beta_1(G) r_1 r_2\right) (r_1+r_2)^4}{\beta_4(G)},\\
\beta_6(G)&=&3 G^2 r_2^5		r_1^7+10 G^2 r_2^6 r_1^6+G r_2^3 \left(3 G r_2^4-148\right) r_1^5+\left(8-488 G r_2^4\right) r_1^4-4r_2 \left(37 G r_2^4-464\right) r_1^3+6000 r_2^2 r_1^2\nonumber\\
&+&1856 r_2^3 r_1+8 r_2^4,
\end{eqnarray}
we write
\begin{eqnarray}
f_G(G)&=&c_0+\frac{c_1}{\sqrt{2\beta_1(G)}}\left\{4\left(r_1+r_2\right)^2-2r_1r_2\beta_1(G)\right.-\sqrt{\frac{2}{3}} \left[-\beta_5(G)-6\left(\left(r_1+r_2\right)^2-\beta_1(G)r_1r_2\right)^2\right.\nonumber\\
&-&\left.2 r_1 r_2 \beta_1(G)\left(16\left(r_1+r_2\right)^2-3\beta_1(G)r_1r_2\right)+2^{2/3}\beta_1(G)\beta_4(G)\right]^{1/2}+\left[-\frac{2\left(1+2^{2/3}\right)\beta_5(G)}{3}\right.\nonumber\\
&-&\frac{8 \sqrt{6} \beta_6(G) (r_1+r_2)^2}{\sqrt{-\beta_5(G)+6 \left(2\left(r_1+r_2\right)^2-\beta_1(G)r_1r_2\right)^2+2 r_1r_2\beta_1(G) \left(16\left(r_1+r_2\right)^2-\beta_1(G)r_1r_ 2\right)-2^{2/3} \beta_1(G) \beta_4(G)}}\nonumber\\
&+&\left.\left.8 \left(2\left(r_1+r_2\right)^2-\beta_1(G)r_1r_2\right)^2+\frac{8 r_1 r_2}{3} \left(16\left(r_1+r_2\right)^2-3\beta_1(G)r_1r_2\right)\right]^{1/2}\right\}^{1/2}.
\end{eqnarray}

The electromagnetic functions are
\begin{eqnarray}
\rho(r)&=&\frac{-3 r^4+r^2 \left(r_1^2-4 r_1 r_2+r_2^2\right)+3 r_1 r_2 \left(r_1^2+r_1	r_2+r_2^2\right)}{\kappa ^2 r_1 r_2 \left(r^2+r_1 r_2\right)^2}+\frac{c_1}{\kappa ^2} \left(\frac{2}{r r_1^3	r_2^3 \left(r^2+r_1 r_2\right)^3} \left(r^6 \left(3 r_1^4+16 r_1^3 r_2+18 r_1^2 r_2^2\right.\right.\right.\nonumber\\
&+&\left.16 r_1 r_2^3+3 r_2^4\right)+8 r^4 r_1 r_2		\left(r_1^2+r_1 r_2+r_2^2\right) \left(r_1^2+5 r_1 r_2+r_2^2\right)+r^2 r_1^2 r_2^2 \left(-3 r_1^4+8 r_1^3r_2-2 r_1^2 r_2^2+8 r_1 r_2^3-3 r_2^4\right)\nonumber\\
&+&\left.\left.8 r_1^4 r_2^4 \left(r_1^2+r_1 r_2+r_2^2\right)\right)+\frac{6 \left(r_1^2-r_2^2\right)^2 }{r_1^{7/2}r_2^{7/2}}\tan ^{-1}\left(\frac{r}{\sqrt{r_1r_2}}\right)\right),\\
L(r)&=&\frac{3 r^6+9 r^4 r_1 r_2+r^2 r_1 r_2 \left(r_1^2+11 r_1 r_2+r_2^2\right)-3 r_1^2 r_2^2 \left(r_1^2+r_1r_2+r_2^2\right)}{\kappa ^2 r_1 r_2 \left(r^2+r_1 r_2\right)^3}\nonumber\\
&+&\frac{c_1}{\kappa^2} \left(\frac{2}{r r_1^3 r_2^3 \left(r^2+r_1r_2\right)^4} \left(r^2 r_1^3 r_2^3 \left(3 r_1^4+8 r_1^3 r_2+26 r_1^2 r_2^2+8
r_1 r_2^3+3 r_2^4\right)-r^8 \left(r_1^2+r_2^2\right) \left(3 r_1^2+8 r_1 r_2+3 r_2^2\right)\right.\right.\nonumber\\
&-&r^4 r_1^2 r_2^2 \left(29 r_1^4+108 r_1^3	r_2+134 r_1^2 r_2^2+108 r_1 r_2^3+29 r_2^4\right)-r^6 r_1 r_2 \left(11r_1^4+32 r_1^3 r_2+26 r_1^2 r_2^2+32 r_1 r_2^3+11 r_2^4\right)\nonumber\\
&-&\left.\left.4 r_1^5 r_2^5 \left(r_1^2+r_1 r_2+r_2^2\right)\right)-\frac{6 \left(r_1^2-r_2^2\right)^2 }{r_1^{7/2} r_2^{7/2}}\tan ^{-1}\left(\frac{r}{\sqrt{r_1r_2}}\right)\right),\\
X(r)&=&\frac{r^4 (r_1+r_2)^2 \left(r^4+6 r^2 r_1 r_2+5 r_1^2 r_2^2\right)}{\kappa ^2 r_1 r_2\left(r^2+r_1 r_2\right)^4}+\frac{8 c_1}{\kappa ^2 r_1^2 r_2^2	\left(r^2+r_1 r_2\right)^4} \left(r^9 \left(2 r_1^2+3 r_1 r_2+2 r_2^2\right)\right.\nonumber\\
&+&4 r^7 r_1 r_2 \left(2 r_1^2+3 r_1 r_2+2 r_2^2\right)-r^5r_1 r_2 \left(6 r_1^4+13 r_1^3 r_2+20 r_1^2 r_2^2+13 r_1 r_2^3+6 r_2^4\right)+2 r^3 r_1^3 r_2^3 \left(3r_1^2+4 r_1 r_2+3 r_2^2\right)\nonumber\\
&+&\left.r r_1^4 r_2^4 \left(r_1^2+r_1 r_2+r_2^2\right)\right).
\end{eqnarray}
The functions $\rho(r)$ and $L(r)$ are well behaved while $X(r)$ diverges in infinity. The modifications came from the nonlinear terms of the $f(G)$ function, since we do not have the presence of $c_0$.
\subsection{Regular black hole with three horizons}
We are not able to find an analytical form of $r(G)$, however, we still can find the corrections to the electromagnetic theory, which are
\begin{eqnarray}
\rho(r)&=&\frac{l^3 r^4+3 l^2 \gamma r^2+\gamma \left(7 r^4-5 \omega r^2+3\delta\right)+l \left(3 r^6-\omega r^4-\delta r^2+6 \gamma^2\right)}{\left(l r^2+\gamma\right)^3 \kappa^2}\nonumber\\
&+&\frac{c_1}{\kappa ^2} \left(\frac{1}{4 l^4 r \gamma^3 \left(l r^2+\gamma\right)^5}\left[75 l^{10} \gamma^2 r^{10}+2 l^9 \gamma \left(33 \delta r^2+175 \gamma^2\right) r^8+l^8 \left(3 \left(5 \delta-6\omega\gamma^2\right) r^4+308\gamma^2\delta r^2+640 \gamma^4\right)
		r^6\right.\right.\nonumber\\
&+&2 l^7 \gamma \left(\left(-3 \delta\omega -11 \gamma^2\right) r^6+7 \left(5 \delta^2-6\omega\gamma^2\right) r^4+288 \gamma^2\delta r^2+273 \gamma^4\right) r^4+105 \gamma^8 r^2+10 l
\gamma^7 \left(49 r^2+3 \omega\right) r^2\nonumber\\
&+&l^2 \gamma^6 \left(896 r^4+140\omega r^2-3 \left(\omega^2+2 \delta\right)\right) r^2+2 l^3 \gamma^5 \left(395 r^6+128 \omega r^4-7 \left(\omega^2+2\delta\right) r^2+3 \left(\delta\omega-7\gamma^2\right)\right) r^2\nonumber\\
&+&l^6\gamma^2 \left(3 \left(\omega^2+2\delta\right) r^8-4	\left(7 \delta\omega+31 \gamma^2 \right) r^6+128 \left(\delta^2-\omega\gamma^2\right) r^4+524 \gamma^2\delta r^2+181 \gamma^4\right) r^2\nonumber\\
&+&l^4 \gamma^4 \left(151r^{10}+500 \omega r^8-128 \left(\omega^2+2 \delta\right) r^6+4 \left(39 \delta\omega-\gamma^2\right) r^4-5 \left(3 \delta^2+22 \gamma^2\omega\right) r^2+64 \gamma^2 \delta\right)\nonumber\\
&+&\left.2 l^5 \gamma^3 \left(49 \omega r^{10}+7 \left(\omega^2-2\delta\right) r^8-32 \left(2\delta\omega+5\gamma^2\right) r^6+3 \left(31 \delta^2+14 \gamma^2\omega \right) r^4-\gamma^2\delta r^2+64 \gamma^4\right)\right]\nonumber\\
&+&\left.\frac{3 \left(25 \gamma^2 l^6+22 \gamma\delta l^5+\left(5\delta^2-6 \gamma^2\omega\right) l^4-2 \gamma \left(\delta\omega-7\gamma^2\right) l^3+\gamma^2 \left(\omega^2+2\delta\right) l^2-10 \gamma^3 \omega l-35 \gamma^4\right) \tan ^{-1}\left(\frac{\sqrt{l} r}{\sqrt{\gamma}}\right)}{4 l^{9/2} \gamma^{7/2}}\right),\nonumber\\\\
L(r)&=&-\frac{\left(3 r^8+\delta r^4-6 \gamma^2 r^2\right) l^2+2\gamma \left(6 r^6+\omega r^4-4 \delta r^2+3 \gamma^2\right) l+\gamma^2 \left(21 r^4-10 \omega r^2+3 \delta\right)}{\left(l r^2+\gamma\right)^4\kappa ^2}\nonumber\\
&-&\frac{c_1}{\kappa ^2} \left(\frac{1}{4l^4 r \gamma^3 \left(l r^2+\gamma\right)^6}\left[75 l^{11}	\gamma^2 r^{12}+l^{10} \gamma \left(66 \delta	r^2+425 \gamma^2\right) r^{10}+l^9 \left(3 \left(5\delta^2-6 \omega \gamma^2\right) r^4+374 \gamma^2\delta r^2+990 \gamma^4\right) r^8\right.\right.\nonumber\\
&+&l^8 \gamma \left(-2 \left(3\delta\omega-5\gamma^2\right) r^6+17 \left(5\delta^2-6\omega\gamma^2\right) r^4+852\gamma^2\delta r^2+1250 \gamma^4\right) r^6+l^7 \gamma^2 \left(3
\left(\omega^2+2\delta\right) r^8\right.\nonumber\\
&-&\left.2 \left(17\delta\omega+11\gamma^2\right) r^6+6 \left(33\delta^2-46\omega\gamma^2\right) r^4+1036 \gamma^2\delta r^2+919 \gamma^4\right) r^4+105 \gamma^9 r^2+5 l \gamma^8 \left(119 r^2+6 \omega\right) r^2\nonumber\\
&+&l^2 \gamma^7 \left(1386 r^4+170\omega r^2-3 \left(\omega^2+2\delta\right)\right) r^2+l^3 \gamma^6 \left(1686 r^6+396 \omega r^4-17 \left(\omega^2+2\delta\right) r^2+6 \left(\delta\omega-7\gamma^2\right)\right) r^2\nonumber\\
&+&l^6\gamma^3 \left(34 \omega r^{10}+17 \left(\omega^2+2\delta\right) r^8+12 \left(3\delta\omega-11\gamma^2\right) r^6+\left(58\delta^2-748\omega\gamma^2\right) r^4+1482\gamma^2\delta r^2-267 \gamma^4\right)r^2\nonumber\\
&+&l^4 \gamma^5 \left(173 r^{10}+1716 \omega r^8-398 \left(\omega^2+2\delta\right) r^6+2 \left(177\omega\delta+121\gamma^2\right) r^4-3\left(5\delta^2+58\omega\gamma^2\right) r^2+32 \gamma^2 \delta\right)\nonumber\\
&+&\left.l^5 \gamma^4 \left(23 r^{12}+214 \omega r^{10}+270 \left(\omega^2+2\delta\right) r^8-4 \left(217\omega\delta+225\gamma^2\right) r^6+5 \left(111\delta^2-\omega\gamma^2\right) r^4-258\gamma^2\delta r^2+64 \gamma^4\right)\right]\nonumber\\
&+&\left.\frac{3 \left(25 \gamma^2 l^6+22 \gamma\delta l^5+\left(5\delta^2-6 \gamma^2\omega\right) l^4-2 \gamma \left(\delta\omega-7\gamma^2\right) l^3+\gamma^2 \left(\omega^2+2\delta\right) l^2-10 \gamma^3 \omega l-35 \gamma^4\right) \tan ^{-1}\left(\frac{\sqrt{l} r}{\sqrt{\gamma}}\right)}{4 l^{9/2} \gamma^{7/2}}\right),\nonumber\\\\
X(r)&=&\frac{r^4 }{\kappa ^2 \left(l r^2+\gamma\right)^4}\left(l^4 r^4+4 l^3 r^2 \gamma+l^2 \left(-r^4\omega-2 r^2 \delta+15 \gamma^2\right)-2 l \gamma \left(r^4+4 r^2
\omega-5 \delta\right)+\gamma^2 \left(5	\omega-14 r^2\right)\right)\nonumber\\
&+&\frac{c_1 r}{\left(l r^2+\gamma\right)^6 \kappa ^2} \left(-8 l^4 \left(r^6-\delta r^2+2	\gamma^2\right) r^6-16 l^3 \gamma \left(3 r^6-\omega r^4-\delta r^2+3 \gamma^2\right) r^4+16 l^2\left(\omega r^{10}-3 \left(\delta\omega+3\gamma^2\right) r^6\right.\right.\nonumber\\
&+&\left.\left(4\delta^2+11\omega\gamma^2\right) r^4-15 \gamma^2\delta r^2+9 \gamma^4\right) r^2+8 \gamma^2 \left(24 r^{10}-30 \omega r^8+8 \left(\omega^2+2\delta\right) r^6-3 \left(2\omega\delta+3\gamma^2\right) r^4+2 \gamma^2	\omega r^2\right.\nonumber\\
&+&\left.\left.\gamma^2\delta\right)+16 l\gamma \left(2 r^{12}+6 \omega r^{10}-6 \left(\omega^2+2\delta\right) r^8+\left(14\delta\omega+9\gamma^2\right) r^6-3 \left(2\delta^2+3\gamma^2\omega\right) r^4+5\gamma^2\delta r^2+\gamma^4\right)\right).
\end{eqnarray}
The conclusions are the same as the previous example.

\end{document}